\documentclass[prc,aps,twocolumn,showpacs,nofootinbib]{revtex4-1}
\usepackage{graphicx}
\newcommand{\be}{\begin{equation}}
\newcommand{\ee}{\end{equation}}
\newcommand{\ba}{\begin{array}}
\newcommand{\ea}{\end{array}}
\newcommand{\gnp}{\gamma n\!\to\!\pi^- p}
\newcommand{\gnn}{\gamma n\!\to\!\pi^0 n}
\newcommand{\gdpp}{\gamma d\!\to\!\pi^- pp}

\newcommand{\mgd}{M_{\gamma d}}

\newcommand{\bfp}{\mbox {\boldmath $p$}}

\newcommand{\crs}{cross section}
\newcommand{\crss}{\crs s}
\newcommand{\ega}{E_{\gamma}}
\newcommand{\sgd}{\sigma_{\gamma d}}
\newcommand{\sgn}{\sigma_{\gamma n}}
\newcommand{\asgn}{\bar{\sigma}_{\gamma n}}

\def\pcomma{$^,$}
\def\pduke{$^1$}
\def\pgwu{$^2$}
\def\pmsstate{$^3$}
\def\pitep{$^4$}
\def\pinfnfr{$^5$}
\def\planl{$^6$}
\def\pjlab{$^7$}
\begin{document}

\title{Amplitude analysis of $\gnp$ data above 1~GeV}

\author{W.~Chen\pduke,
	H.~Gao\pduke,
	W.~J.~Briscoe\pgwu,
	D.~Dutta\pmsstate,
	A.~E.~Kudryavtsev\pitep\pcomma\pgwu,
	M.~Mirazita\pinfnfr,
	M.~W.~Paris\planl,
	P.~Rossi\pinfnfr, 
	S.~Stepanyan\pjlab,
	I.~I.~Strakovsky\pgwu,
	V.~E.~Tarasov\pitep,
	R.~L.~Workman\pgwu\\ 
\vspace*{0.3in}}
\affiliation{
\pduke Duke University, Durham, NC 27708, USA}
\affiliation{
\pgwu The George Washington University, Washington, DC 20052, USA}
\affiliation{
\pmsstate Mississippi State University, Mississippi State, MS 39762, 
	USA}
\affiliation{
\pitep Institute of Theoretical and Experimental Physics, Moscow, 
	117259 Russia}
\affiliation{
\pinfnfr INFN, Laboratori Nazionali di Frascati, 00044 Frascati, Italy}
\affiliation{
\planl Theory Division, Los Alamos National Laboratory, Los Alamos, NM 
	87545, USA}
\affiliation{
\pjlab Thomas Jefferson National Accelerator Facility, Newport News, VA 
	23606, USA\\}

\date{\today}

\begin{abstract}
We report a new extraction of nucleon resonance couplings using
$\pi^-$ photoprodution cross sections on the neutron. The world
database for the process $\gamma n\to\pi^-p$ above 1~GeV has
quadrupled with the addition of new differential cross sections
from the CEBAF Large Acceptance Spectrometer (CLAS) at Jefferson
Lab in Hall~B. Differential cross sections from CLAS have been 
improved with a new final-state interaction determination
using a diagrammatic technique taking into account the $NN$ and
$\pi$N final-state interaction amplitudes. Resonance couplings 
have been extracted and compared to previous determinations. With 
the addition of these new \crss\, significant changes are seen in 
the high-energy behavior of the SAID \crss\ and amplitudes.
\end{abstract}

\pacs{13.60.Le, 24.85.+p, 25.10.+s, 25.20.-x}
\maketitle

\section{Introduction}
\label{sec:intro}

High-precision data and new analysis techniques for $\gamma N\to\pi 
N$ are beginning to have a transformative impact on our understanding 
of the $N$ and $\Delta$ resonance spectrum. With the arrival of new 
and improved measurements of single- and double-polarization 
quantities, fits have become highly constrained. As a result, some 
multipole amplitudes and their underlying resonant components have 
changed significantly. This is particularly true for the 
neutron-target sector, where, until recently, there were few data 
on which to base fits and from which to extract $n\gamma$ photo-decay 
amplitudes.  

The radiative decay width of the neutral states may be extracted from
$\pi^-$ and $\pi^0$ photoproduction off a neutron, which involves a 
bound neutron target (typically the deuteron) and requires the use of 
a model-dependent nuclear correction.  As a result, our knowledge of 
neutral resonance decays is less precise compared to the charged ones.

The existing database contains mainly $\gnp$ differential cross sections.  
Many of these are old bremsstrahlung measurements with limited angular 
coverage and broad energy binning.  In several cases, the systematic 
uncertainties have not been given.  At lower energies, there are 
data sets for the inverse $\pi^-$ photoproduction reaction: 
$\pi^-p\to\gamma n$.  
This process is free from complications associated with a deuteron 
target.  However, the disadvantage of using this reaction is the high 
background from  the 5 to 500 times larger \crs\ for 
$\pi^-p\to\pi^0n\to\gamma\gamma n$.

Here we explore the effect of adding CLAS differential \crss\ for 
$\gnp$, extracted from $\gdpp$~\cite{CLAS}, to the full SAID 
database. Measurements extend from 1.05 to 3.5~GeV in the photon 
energy.  The present \crs\ set has quadrupled the world database for 
$\gnp$ above 1~GeV, which allows for fits covering the region up to
2.7~GeV. We will show that these new data require large adjustments 
of our fits.

In the next section, Sec.~\ref{sec:DB}, we will give a brief overview 
of the available experimental data. A discussion of the final-state
interaction (FSI) calculations is given in Sec.~\ref{sec:FSI}. The 
new CLAS data are compared with fits and older measurements in Sec.
\ref{sec:results}. In Sec.~\ref{sec:fit}, we discuss the fits and the 
extraction of resonance parameters. Finally, in Sec.~\ref{sec:conc}, 
we summarize our findings and discuss the potential impact of future
measurements and partial-wave analysis (PWA).

\section{Data Set}
\label{sec:DB}

Due to lack of neutron targets, the database for the reactions $\gnn$ 
and $\gnp$ is small compared to single-pion photoproduction reactions 
using proton targets, $\gamma p\to\pi^+n$ and $\gamma p\to\pi^0p$.
Previous $\gamma N\to\pi N$ measurements are available in the SAID 
database~\cite{world_data}.
  
Only $364$ data-points are available for $\gnn$ below 2.7~GeV.  For 
$\gnp$, the situation is especially dire in the photon energy range 
above 1~GeV. There are only 360 data points, half of which come from 
polarized measurements. Below 1~GeV, there are significant numbers of 
$\gnp$ data, coming mainly from Meson Factories (LAMPF, TRIUMF, and 
PSI) via inverse pion photoproduction $\pi^-p\to\gamma n$.  Overall, 
there are $2093$ data points, 17\% of which are from polarized 
measurements. Some differential \crss\ for the $\pi^-p\to\gamma n$ 
have been measured at BNL AGS, using the Crystal Ball multiphoton 
spectrometer. Measurements were made at 18 pion momenta from 238 to 
748~MeV/$c$, corresponding to E$_\gamma$ from 285 to 769~MeV~\cite{aziz}.  
These data have been used to evaluate neutron multipoles in the vicinity 
of the $N(1440)1/2^+$ resonance.  

We have recently considered the effect of the beam-asymmetry data 
($\Sigma$) of $\vec{\gamma}n\to\pi^-p$~\cite{graal1} and $\vec{\gamma}n
\to\pi^0n$~\cite{graal2} from GRAAL on our fits to neutron-target data
\cite{sn11}. These include $216$ $\Sigma$ measurements of $\pi^0n$ 
covering E$_\gamma$=703--1475~MeV and $\theta$=53--164$^\circ$ plus 
$99$ $\Sigma$ measurements of $\pi^-p$ for E$_\gamma$=753--1439~MeV 
and $\theta$=33--163$^\circ$. Predictions for $\gnn$ were qualitatively 
different from the measurements over a wide angular range above a 
center-of-mass (CM) energy of 1650~MeV.

In 2009, the CLAS Collaboration at Jefferson Lab reported a detailed 
study of the reaction $\gnp$ using a high statistics photoproduction 
experiment on deuterium~\cite{CLAS}. This data set added $855$ 
differential \crss\ between 1.05 and 3.5~GeV, and pion production 
center-of-mass angles between 32$^\circ$ and 157$^\circ$, to the 
existing data base.  The 
overall systematic uncertainty varies between 5.8\%, at the lowest 
photon energy, and up to 9.4\% at the highest photon energy.  Details 
of the data processing and analysis can be found in Ref.~\cite{Wei}. 
An improvement in the FSI has been made since the original publication
\cite{CLAS}.

Chen, \textit{et al.}~\cite{CLAS} estimated FSI corrections according 
to the Glauber formulation~\cite{trans} and this correction was found 
to be about 20\%. The uncertainty of the Glauber calculation for the 
FSI correction was estimated to be 5\% in  Ref.~\cite{zhu,zhu1}. To 
study the model uncertainty in calculating the FSI correction, another 
calculation using the approach of Ref.~\cite{laget} was adopted. Both 
methods agreed within 10\%. A 10\%  systematic uncertainty to the 
differential \crs\ was assigned for the FSI correction~\cite{CLAS}.  

In a further study of the FSI corrections for the $\gamma n\to\pi^-p$ 
\crs\ determination from the deuteron data, we used a diagrammatic 
technique~\cite{PRC2011}, including the fact that CLAS does not detect 
protons with momenta less than 200~MeV/$c$. A short description of the 
FSI formalism is given in Sec.~\ref{sec:FSI}.


\section{FSI calculations}
\label{sec:FSI}

\subsection{Amplitudes}
\label{sec:amp}
\vspace{-3mm}

\begin{figure}
\begin{center}
\includegraphics[height=2.3cm, keepaspectratio]{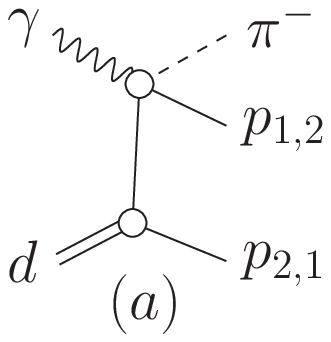}~
\includegraphics[height=2.3cm, keepaspectratio]{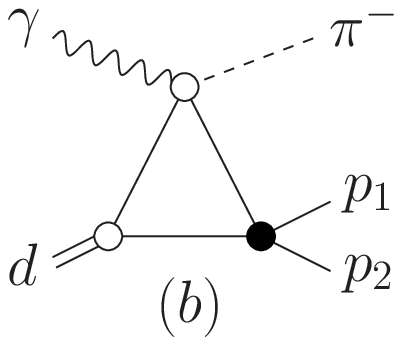}~~
\includegraphics[height=2.3cm, keepaspectratio]{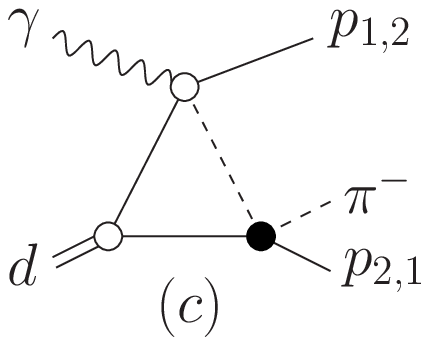}
\end{center}
\caption{Feynman diagrams for the leading components of the
        $\gdpp$ amplitude. (a) IA, (b) $pp$-FSI, and (c)
        $\pi$N-FSI. Filled black circles show FSI vertices.
        Wavy, dashed, solid, and double lines correspond to
        the photons, pions, nucleons, and deuterons,
        respectively.} \label{fig:g1}
\end{figure}

Calculations of the $\gdpp\,$ differential \crss\, with the FSI
taken into account, were done in a model represented by the
diagrams in Fig.~\ref{fig:g1}. These diagrams correspond to the
IA [Fig.~\ref{fig:g1}(a)], $pp$-FSI [Fig.~\ref{fig:g1}(b)], and
$\pi$N-FSI [Fig.~\ref{fig:g1}(c)] amplitudes, denoted by $M_a$,
$M_b$, and $M_c$, respectively. The resulting amplitude $\mgd$
reads
$$
        \mgd=M_a+M_b+M_c,
$$
\be
        M_{a,c}\!=M^{(1)}_{a,c}+M^{(2)}_{a,c}.
\label{amp1}\ee

IA and $\pi$N-FSI diagrams [Figs.~\ref{fig:g1}(a),(c)] include
also the cross-terms  
between the final protons. The terms in Eq.~(\ref{amp1})
depend on the elementary $\gamma N\!\to\!\pi N$ amplitudes and
deuteron wave function (DWF). The terms $M_b$ and $M_c$ depend
also on the $N\!N\!\to\!N\!N$ and $\pi N\!\to\!\pi N$ amplitudes,
respectively.

The $\gamma N\!\to\!\pi N$ amplitudes were expressed through
four independent Chew-Goldberger-Low-Nambu (CGLN) amplitudes
\cite{CGLN} $F_{1-4}$, which were generated by the SAID code,
using the George Washington University (GW) Data Analysis Center
(DAC) pion photoproduction multipoles~\cite{SAID02,pr_PWA}. The
$N\!N$- and $\pi N$-scattering amplitudes were calculated, using
the results of GW $N\!N$~\cite{NN_PWA} and $\pi N$~\cite{piN_PWA}
PWAs. The DWF was taken from the Bonn potential (full model)
\cite{BonnCD}.  The elementary amplitudes are dependent on the
momenta of the external and intermediate particles in Fig.
\ref{fig:g1}. Thus, Fermi motion is taken into account in the
$\gdpp$ amplitude $\mgd$.
Details of calculations of the amplitudes $M_{a,b,c}$ in
Eq.~(\ref{amp1}) are given in Ref.~\cite{PRC2011}.

\vspace{-5mm}
\subsection{FSI Correction}
\label{sec:cor}
\vspace{-3mm}

We extract the $\gnp\,$ \crs\ from the deuteron data in the
quasi-free (QF) kinematical region of the $\gdpp\,$ reaction with
fast and slow protons $p_1$ and $p_2$, respectively, where the
$\gdpp$ \crs\ is dominated by the IA amplitude $M^{(1)}_a$, i.e.,
$\mgd\approx M^{(1)}_a$, while the cross term $M^{(2)}_a$ and the
FSI amplitudes $M_{b,c}$ are relatively small. This consideration
is addressed in the analysis of the CLAS~\cite{Wei,CLAS} data for
the reaction $\gdpp$ with kinematical cuts $|\bfp_2|<200~MeV/c<|\bfp_1|$,
corresponding to the CLAS experimental conditions.

In the QF approximation, the $\gdpp$ and $\gnp\,$ differential \crss\
for unpolarized particles are related to each other in a known
way~\cite{BL}. 
\be
   \frac{d\sgd^{QF}}{d\bfp_2\,d\Omega}=
   \frac{\ega^{\,\prime}}{\ega}\,\rho(p_2)\,\frac{d\sgn}{d\Omega}.
\label{ne1}\ee
Here: $\Omega$ is the solid angle of relative motion in the $\pi^-p_1$
system; $\ega$ and $\ega^{\,\prime}=(1+\beta\cos\theta_2)\ega$ are the
photon laboratory energies for the reactions $\gdpp$ and $\gnp\,$,
respectively; $\beta=p_2/E_2$ ($\theta_2$) is the laboratory velocity
(polar angle) of spectator proton $p_2\,$; $\rho(p)$ is the square of
DWF and $\int\!\rho(p)\,d\bfp=1$. Let $d\sgd^{QF}\!/d\Omega$ and
$d\sgd\!/d\Omega$ be the deuteron \crs, integrated over $\bfp_2$ in
a small region $|\bfp_2|<p_{\,max}$ and obtained with the amplitudes
$\mgd\!=M^{(1)}_a$ and $M_a\!+M_b\!+M_c$, respectively. Then, from
Eq.~(\ref{ne1}) (see details in Ref.~\cite{PRC2011}) we obtain
$$
  \frac{d\asgn^{\,exp}}{d\Omega}(\bar\ega,\theta)= c^{-1}
  R^{-1}\!(\ega,\theta)\frac{d\sgd^{exp}}{d\Omega}(\ega,\theta),
$$

\vspace{-5mm}
\be
   c=\mkern-22mu \int\limits_{p\,<p_{max}}\mkern-22mu \rho(\bfp)\,d\bfp,
~~~~~R(\ega,\theta)=\frac{d\sgd\!/d\Omega}{d\sgd^{QF}\!/d\Omega},
\label{ne2}\ee
were $d\asgn^{\,exp}/d\Omega$ is the neutron \crs, extracted from
the deuteron data. Here: $\theta$ is the polar angle of the outgoing
pion in the $\pi^- p_1$ frame; $c=c(p_{max})\le 1$ is the ``effective
number" of neutrons with momenta $p<p_{max}$ in the deuteron; $R$ is
the correction factor for FSI effects as well as for the ``suppressed"
amplitude $M^{(2)}_a$. The factor $R$ depends on $\ega$ and $\theta$
as well as on the kinematical cuts applied.

The neutron \crs\ $d\asgn/d\Omega$~(\ref{ne2}) is averaged over
the photon energy $\ega'\sim\ega$, and $\bar\ega$ is some
``effective" $\ega'$ value in the range $\ega(1\pm\beta)$. For small
values of $p_{max}$ we have $\beta\ll 1$ and $\bar\ega\approx\ega$.

\begin{figure*}[th]
\centerline{
\includegraphics[height=0.32\textwidth, angle=90]{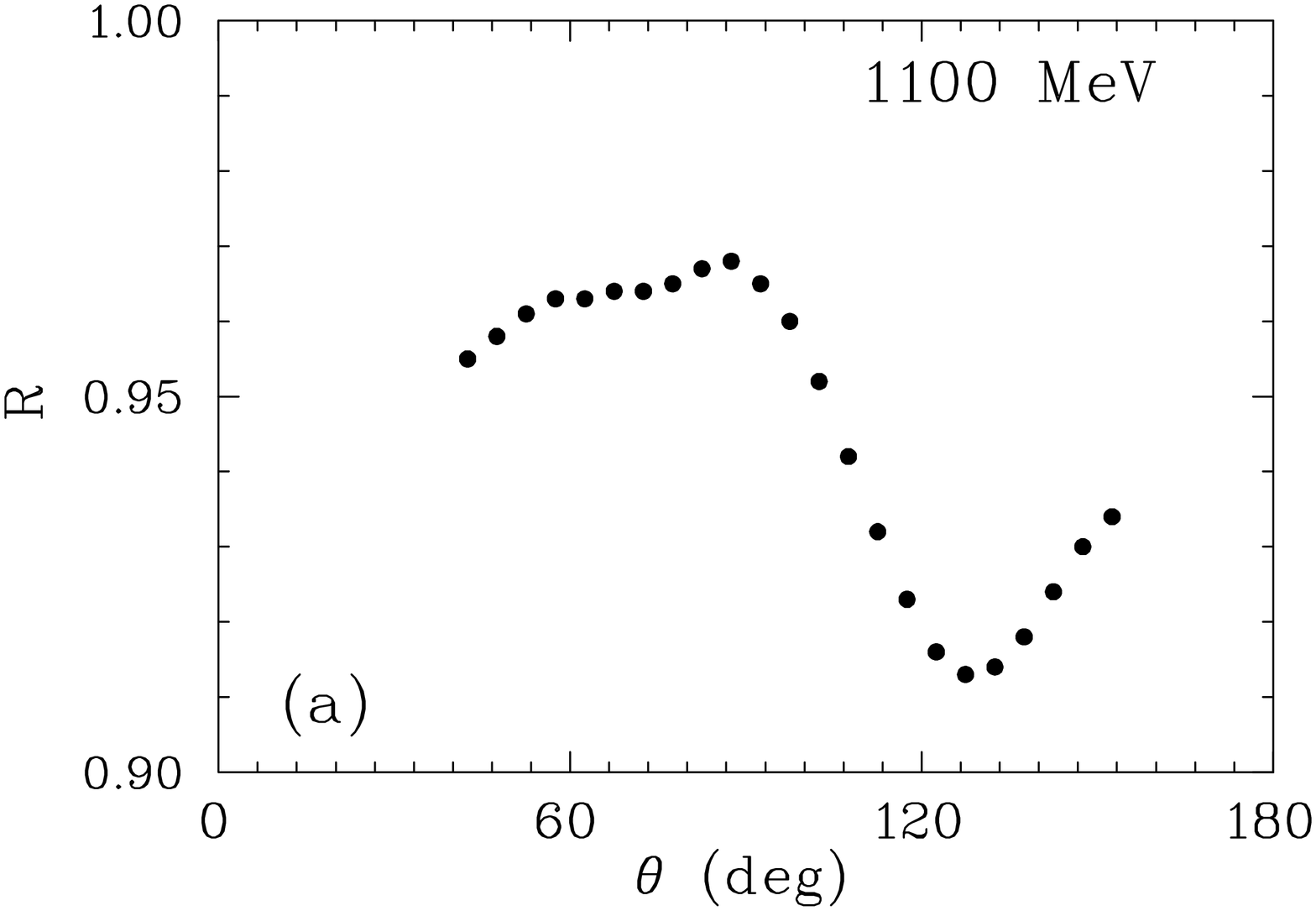}\hfill
\includegraphics[height=0.32\textwidth, angle=90]{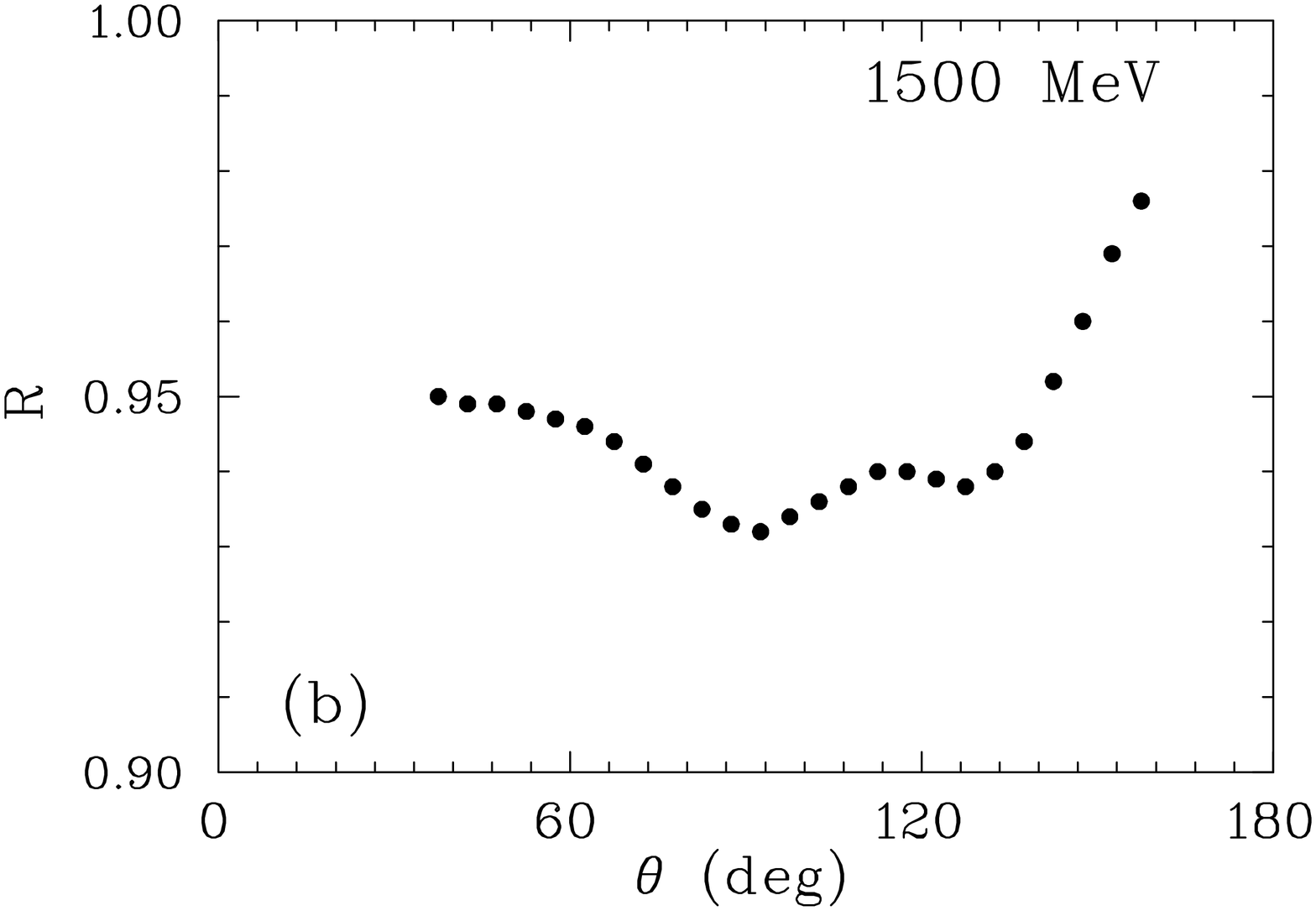}\hfill
\includegraphics[height=0.32\textwidth, angle=90]{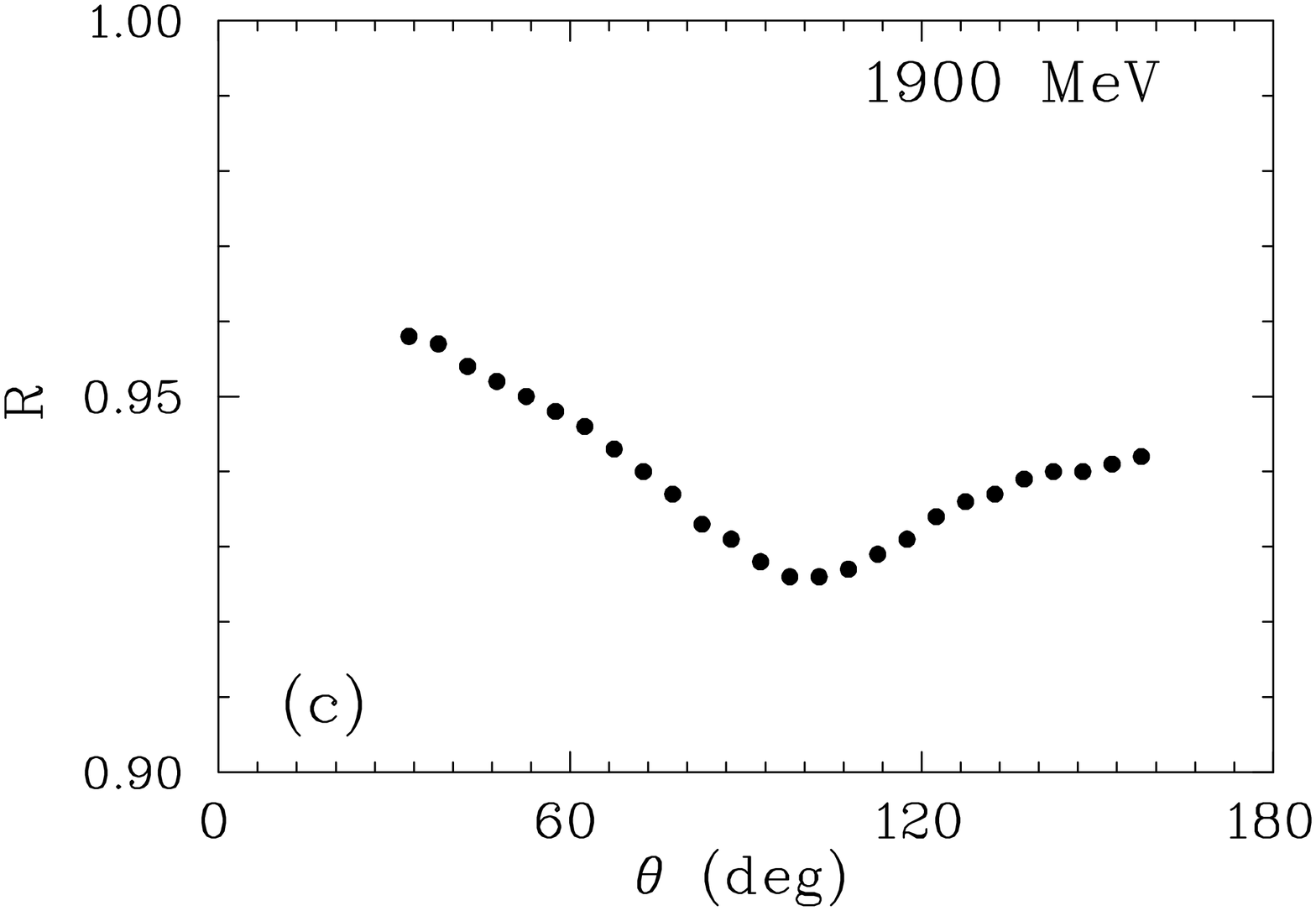} }
\centerline{
\includegraphics[height=0.32\textwidth, angle=90]{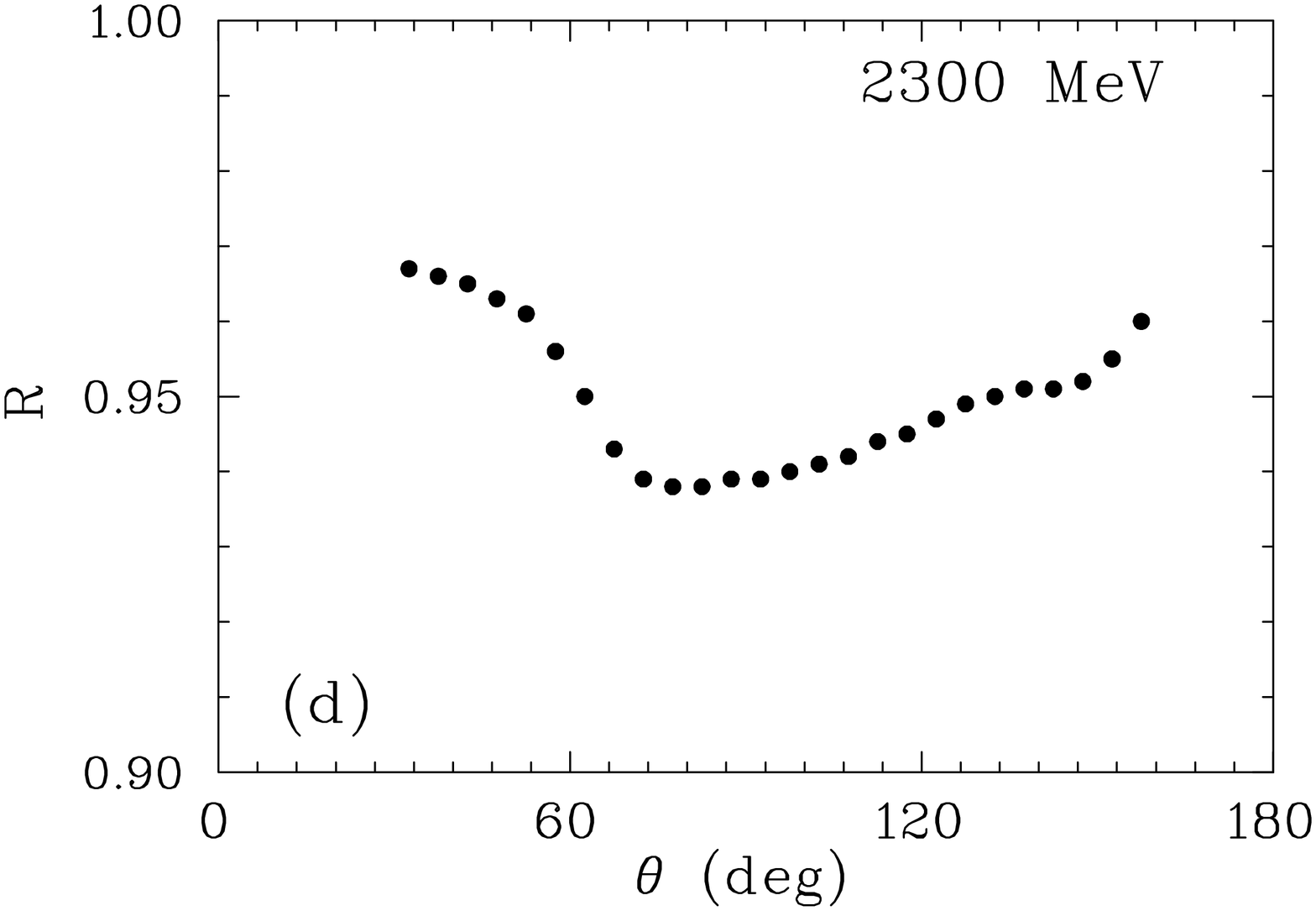}\hfill
\includegraphics[height=0.32\textwidth, angle=90]{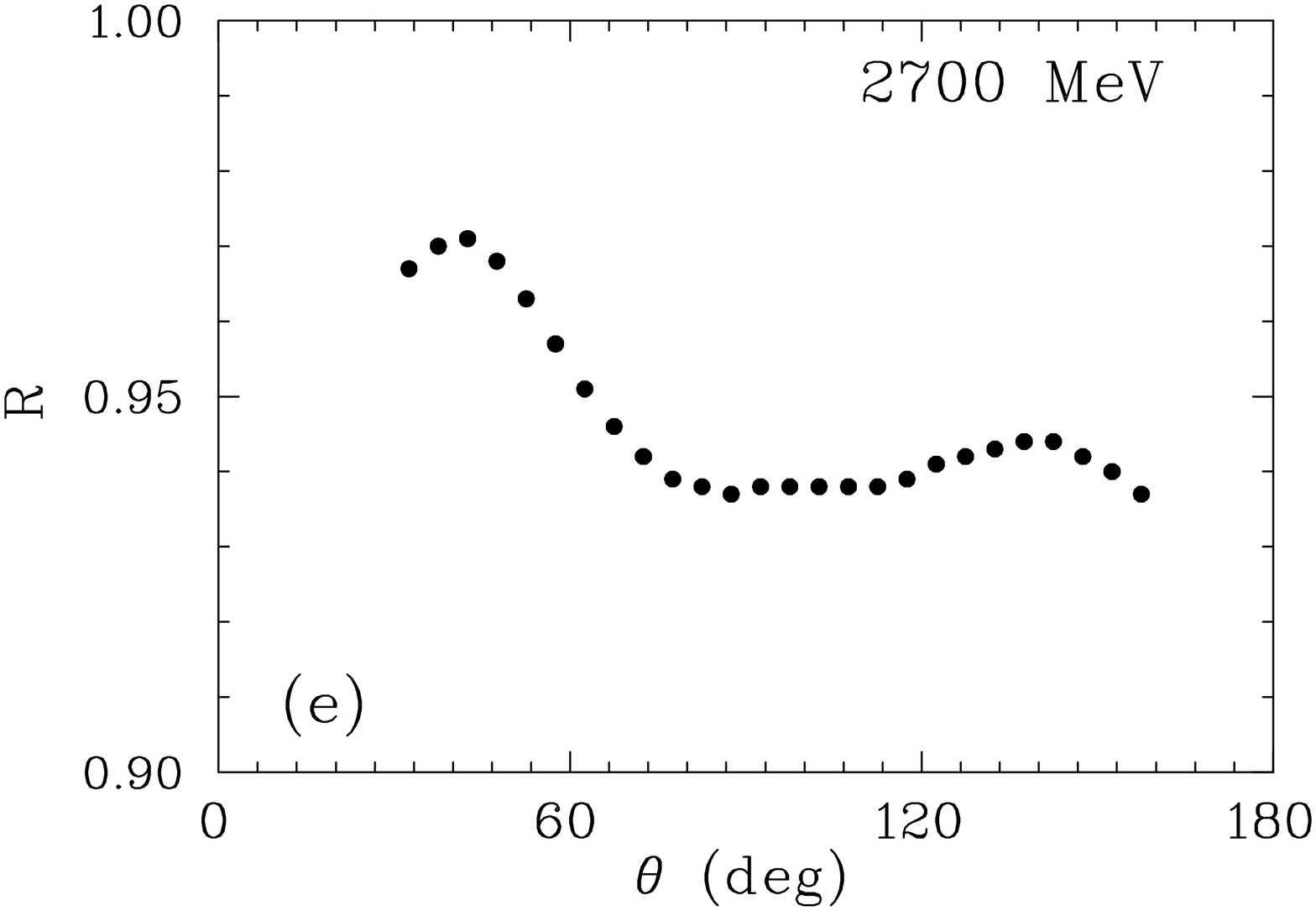}}
\caption{FSI correction factor $R$ for $\gnp$ as a function
        of $\theta$, where $\theta$ is the production angle of
        $\pi^-$ in the CM frame.  The present calculations
        (solid circles) are shown for five energies:
        (a) $E_\gamma$ = 1100~MeV,
        (b)              1500~MeV,
        (c)              1900~MeV,
        (d)              2300~MeV, and
        (e)              2700~MeV. There are no uncertaintes
        given. \label{fig:g2}}
\end{figure*}

We applied FSI corrections~\cite{PRC2011} dependent on the E$_\gamma$
and $\theta$. As an illustration, Fig.~\ref{fig:g2} shows the FSI
correction factor $R$ for the present $\gnp$ differential \crss\ as a
function of the pion production angle in the CM frame for different
energies over the range of the CLAS experiment.  Overall, the FSI
correction factor $R< 1$, while the effect, i.e., the $(1-R)$ value,
is less than 10\% and the behavior is very smooth vs. pion production 
angle.

The contribution of FSI calculations~\cite{PRC2011} to the overall
systematics is estimated to be 2\% (3\%) below (above) 1800~MeV.
Above 2700~MeV, our estimation of systematic uncertainty due to the
FSI calculations is 5\%.  Then we added FSI systematics to the overall 
experimental systematics in quadrature.

\section{Results}
\label{sec:results}

Since the CLAS results for the $\gnp$ differential \crss\ consist 
of $855$ experimental points, they are not tabulated in this or 
the previous~\cite{CLAS} publication, but are available in the 
SAID database~\cite{world_data} along with their uncertainties and 
the energy binning.  

\begin{figure*}[th]
\centerline{
\includegraphics[height=0.32\textwidth, angle=90]{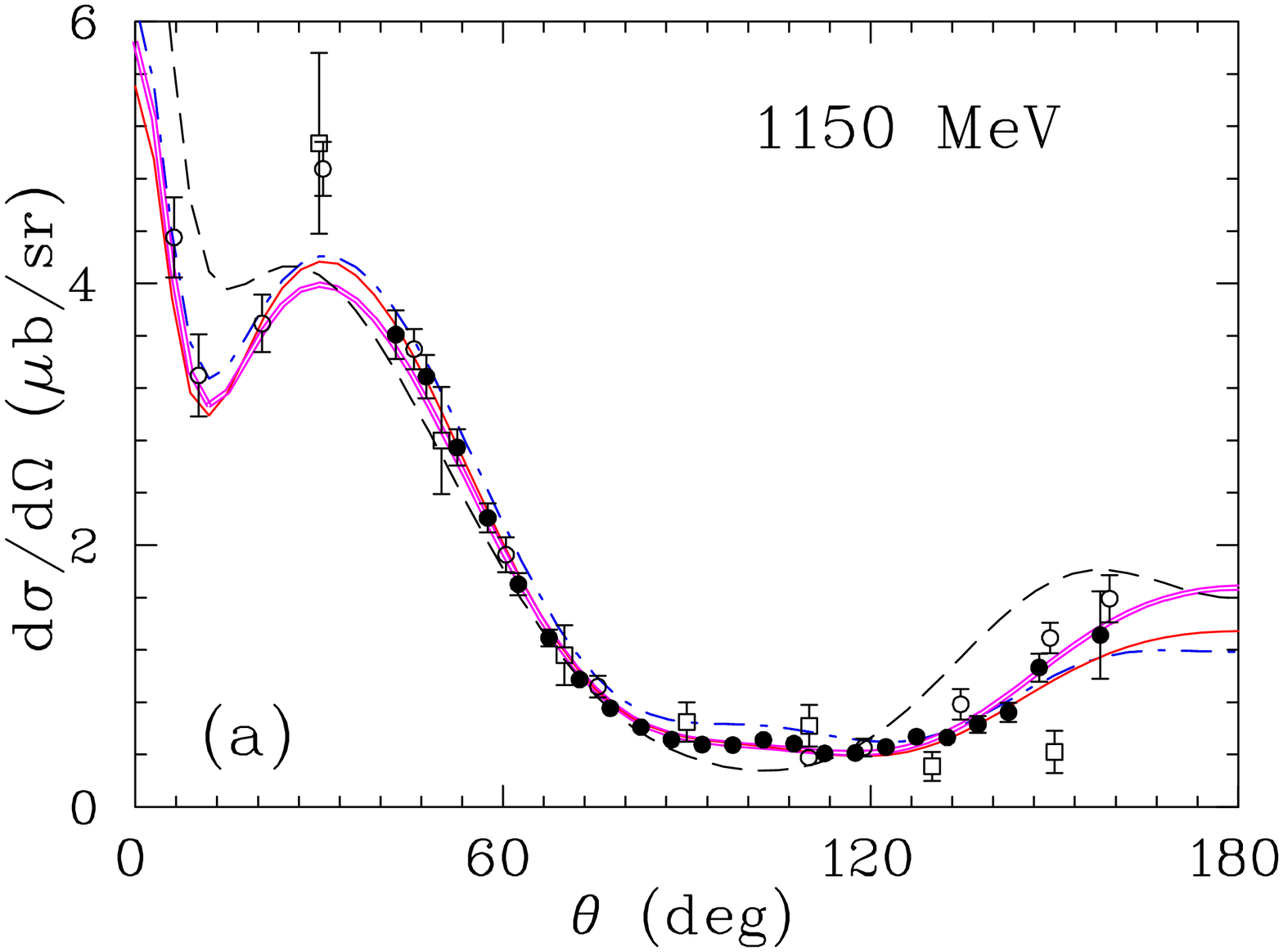}\hfill
\includegraphics[height=0.32\textwidth, angle=90]{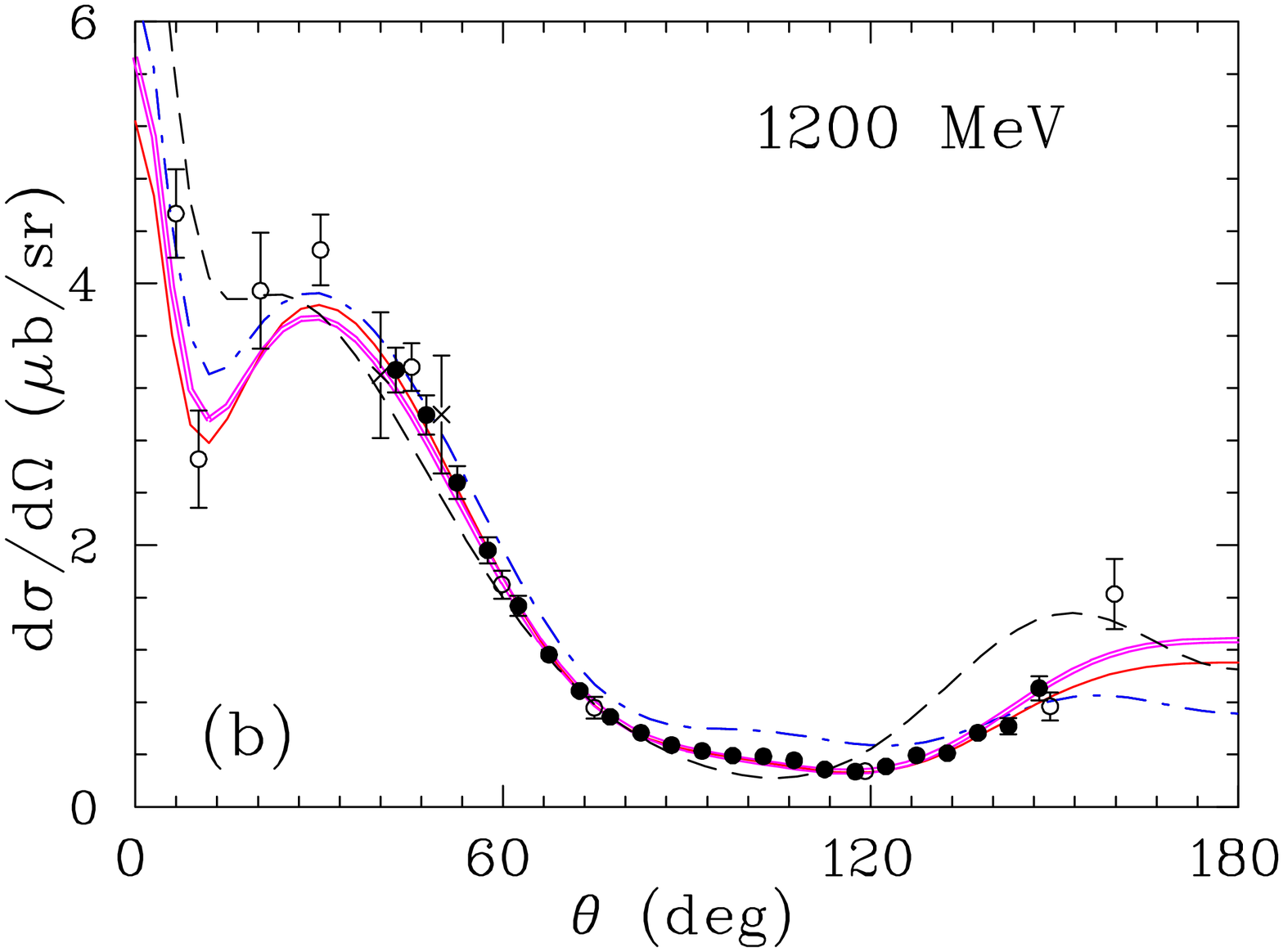}\hfill
\includegraphics[height=0.32\textwidth, angle=90]{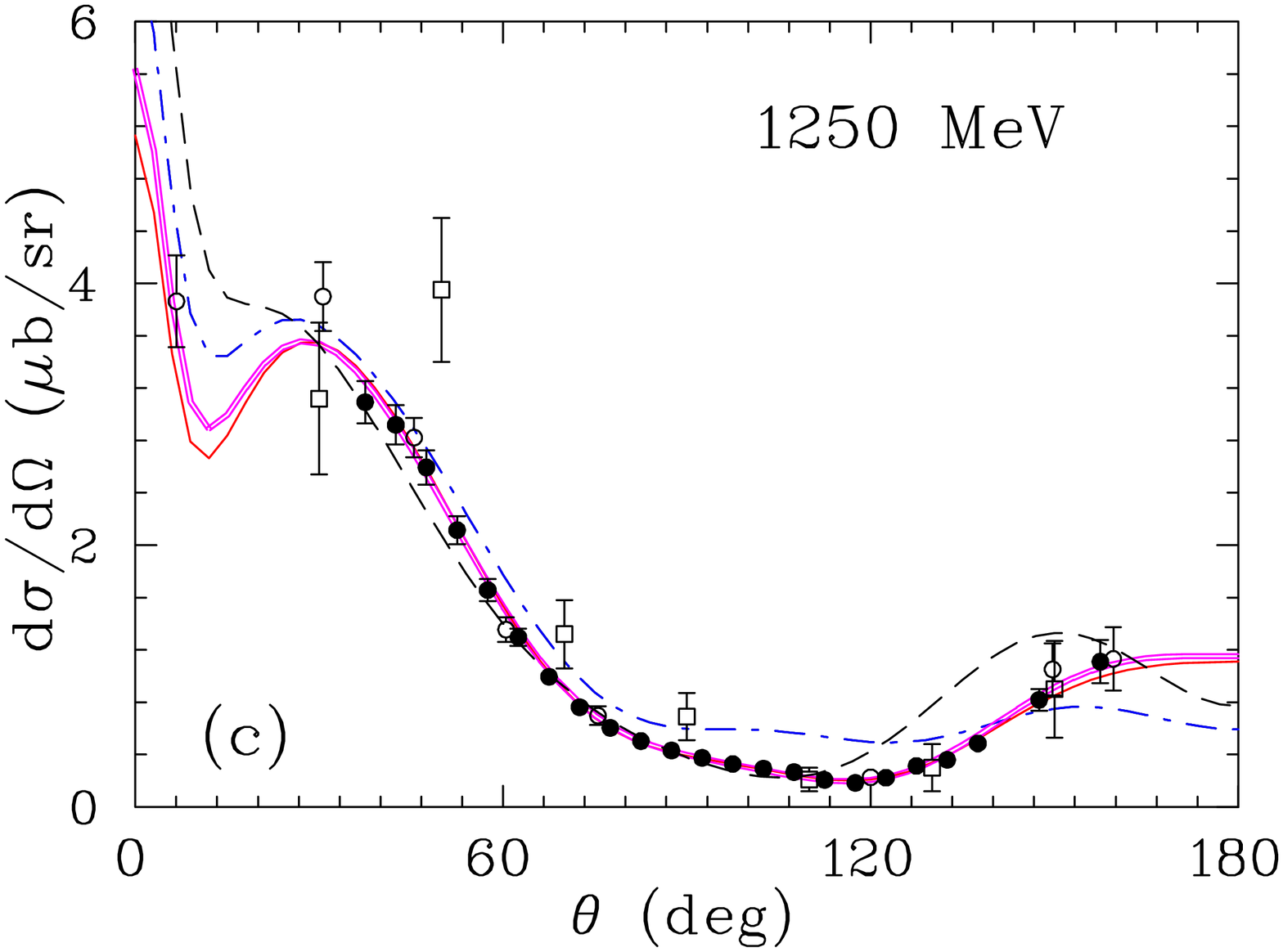}}
\centerline{
\includegraphics[height=0.32\textwidth, angle=90]{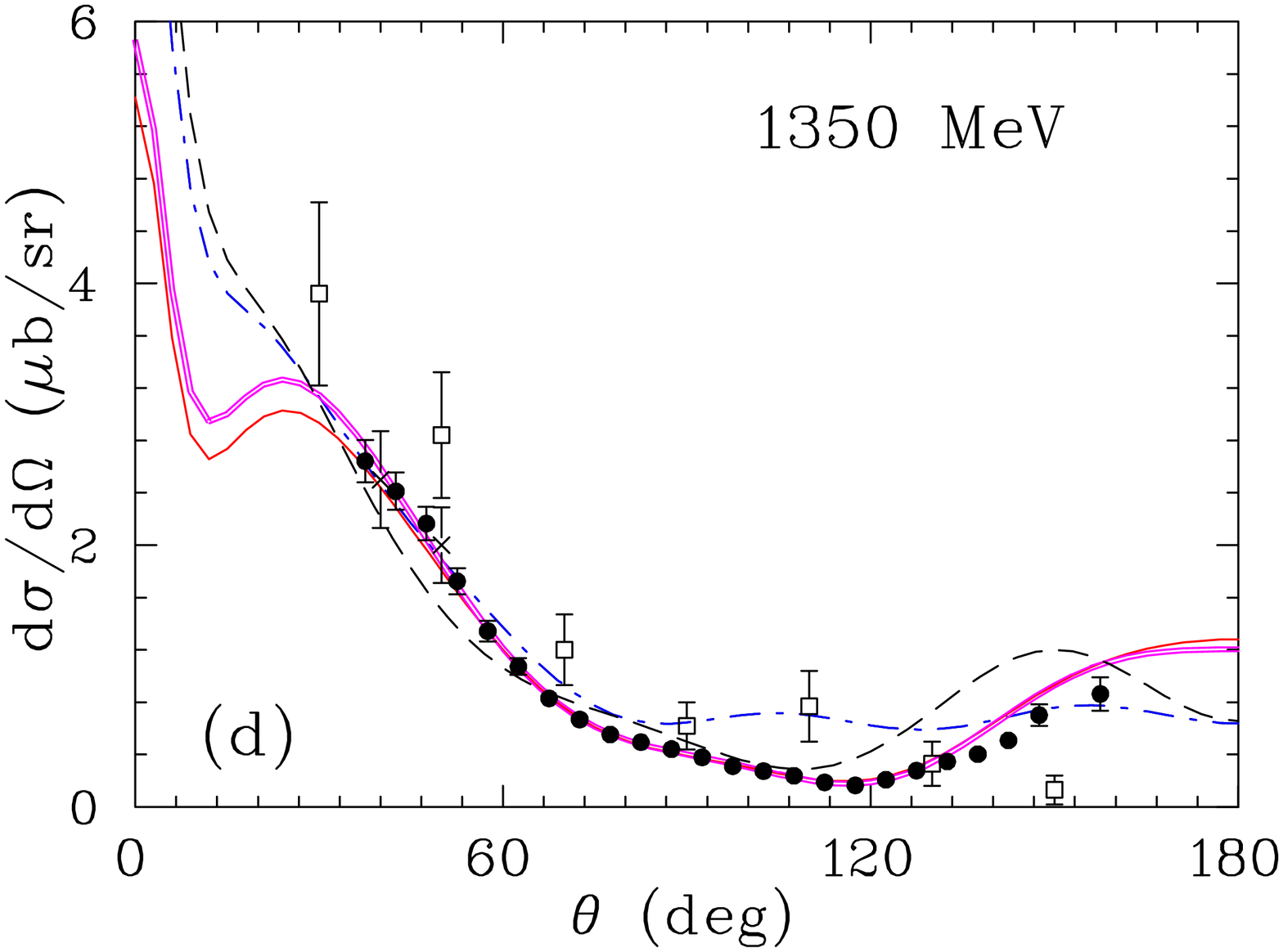}\hfill
\includegraphics[height=0.32\textwidth, angle=90]{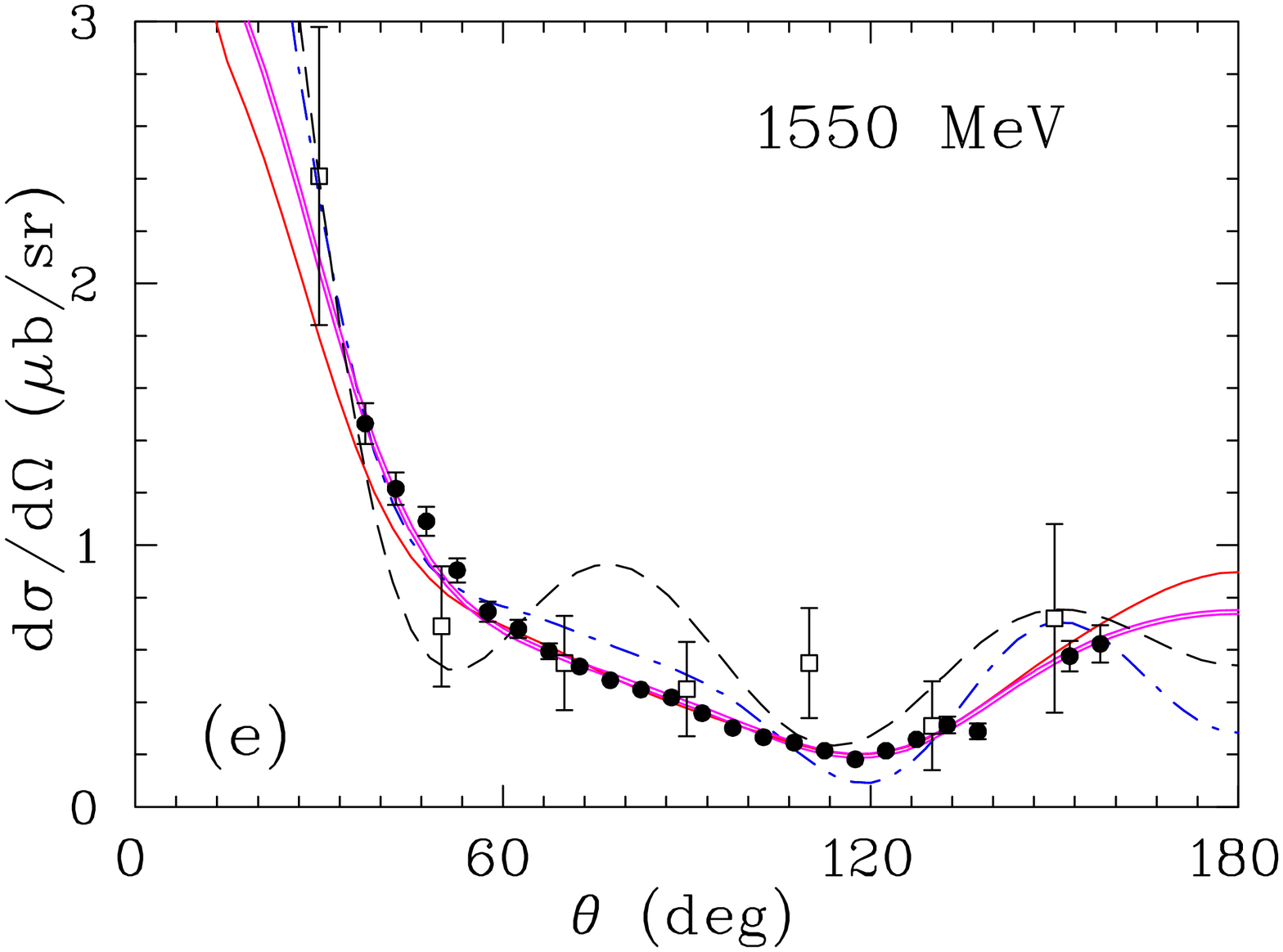}\hfill
\includegraphics[height=0.32\textwidth, angle=90]{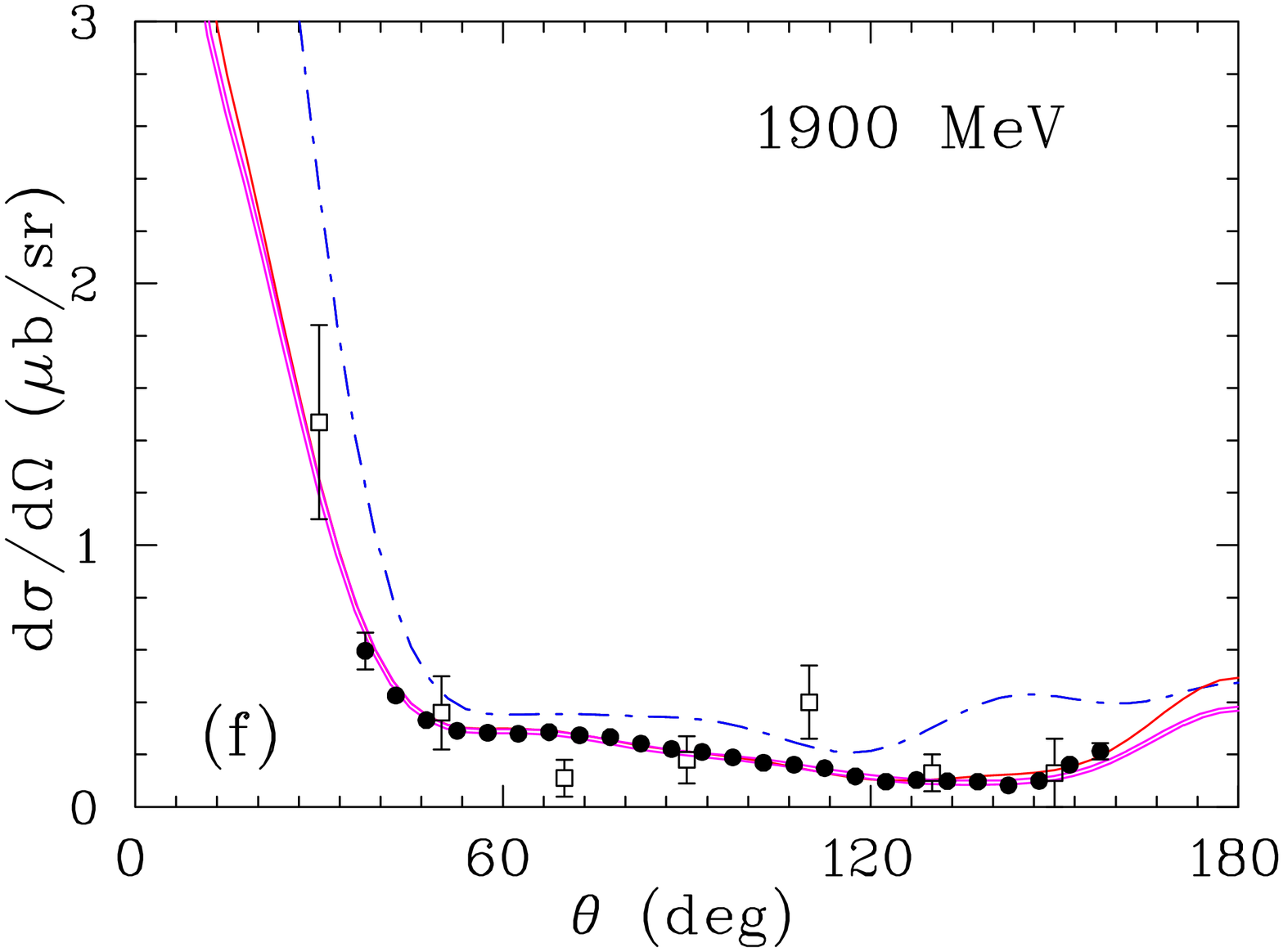}}
\caption{(Color online) Differential \crss\ for $\gnp$ as a 
	function of $\theta$, where $\theta$ is the production 
	angle of $\pi^-$ in the CM frame. The present data 
	(solid circles) are shown for six energy bins:
        (a) $E_\gamma$ = 1150~MeV,
        (b)              1200~MeV,
        (c)              1250~MeV, 
        (d)              1350~MeV,
        (e)              1550~MeV, and
        (f)              1900~MeV.
        Previous data are shown for experiments at SLAC
	\protect\cite{sf74} (open circles); DESY
	\protect\cite{be73} (open squares), and Yerevan
	\protect\cite{ab80} (crosses). Plotted uncertainties 
	are statistical only. Solid (dash-dotted) lines 
	correspond to the GB12 (SN11~\protect\cite{sn11}) 
	solution.  Thick solid (dashed) lines give GZ12 
	solution (MAID07~\protect\cite{maid}, which 
	terminates at $W$ = 2~GeV). \label{fig:g3}}
\end{figure*}

Specific examples of agreement with previous measurements are 
displayed in Fig.~\ref{fig:g3}, where we compare differential 
\crss\ obtained here with those from SLAC~\cite{sf74}, DESY
\cite{be73}, and Yerevan~\cite{ab80}, at energies common to 
those experiments.  Previous measurements used a modified Glauber
approach and the procedure of unfolding the Fermi motion of the
neutron target.  The CLAS data and the results from SLAC, DESY, 
and Yerevan appear to agree well at these energies.  Unfortunately, 
there are no measurements for $\pi^-p\to\gamma n$ to compare
at these energies.

While agreement with previous measurements is generally good, the 
new data extend to higher energies with more complete angular 
coverage and are more constraining in the PWA, as is evident in 
Fig.~\ref{fig:g3}.  

\begin{figure*}[th]
\includegraphics[height=0.913\textwidth, angle=90]{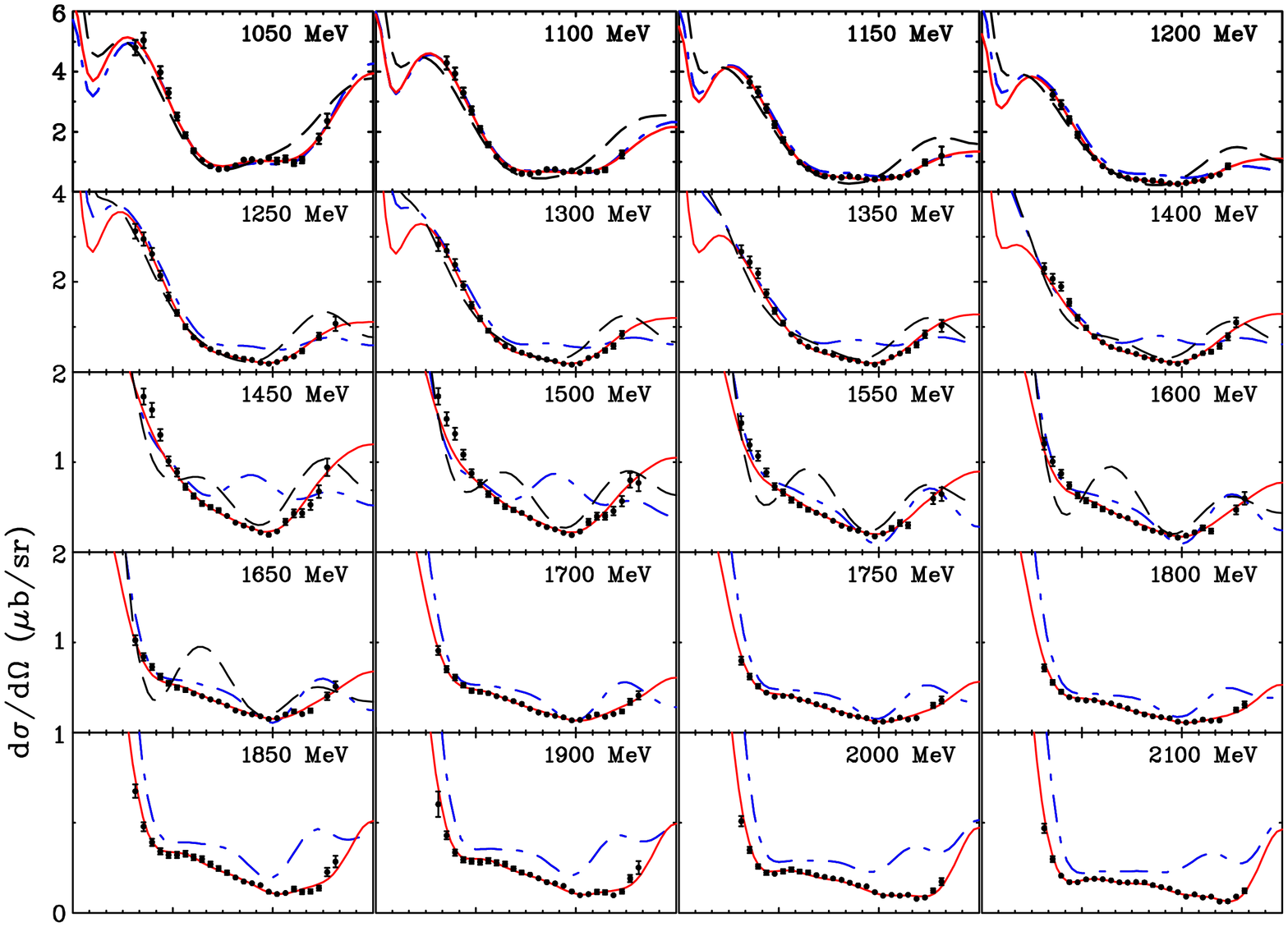}\\
\includegraphics[height=0.9\textwidth, angle=90]{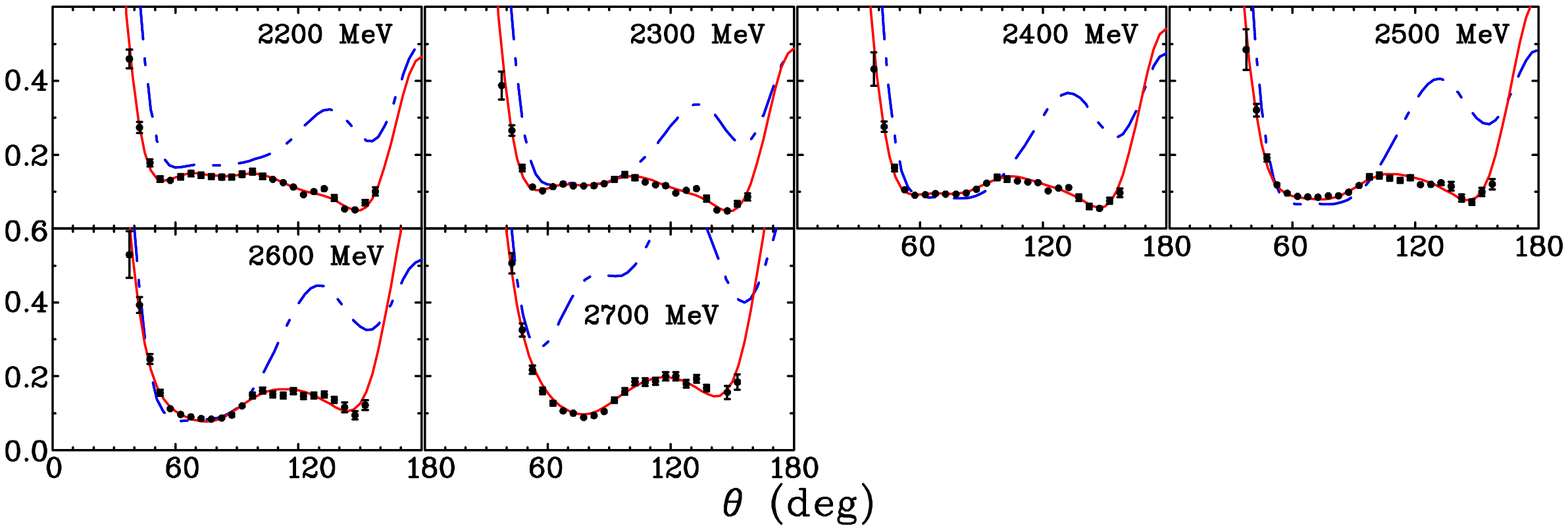}
\caption{(Color online) The differential \crs\ for $\gnp$ below 
	E$_\gamma$ = 2.7~GeV versus pion CM angle.  Solid 
	(dash-dotted) lines correspond to the GB12 (SN11
	\protect\cite{sn11}) solution. Dashed lines give 
	the MAID07 \protect\cite{maid} predictions.  
	Experimental data are from the current (filled 
	circles). Plotted uncertainties are statistical. 
	\label{fig:g4}}
\end{figure*}

A more complete comparison of the CLAS data with fits and 
predictions is given in Fig.~\ref{fig:g4}. It is interesting to 
note that the data appear to have fewer angular structures than 
the earlier fits. 

\section{Amplitude Analysis of Data}
\label{sec:fit}

We have included the new \crss\ from the CLAS experiment in a number 
of multipole analyses covering incident photon energies up to 
2.7~GeV, using the full SAID database, in order to gauge the influence 
of these measurements, as well as their compatibility with previous 
measurements.  

\begin{table}[th]
\caption{$\chi^2$ comparison of fits to pion photoproduction
        data up to 2.7~GeV.  Results are shown for six different 
	SAID solutions (current GB12 and GZ12 with previous SN11
	\protect\cite{sn11}, SP09~\protect\cite{du1}, 
	FA06~\protect\cite{pr_PWA}, SM02~\protect\cite{SAID02}, 
	and SM95~\protect\cite{sm95}). \label{tab:tbl1}}
\vspace{2mm}
\begin{tabular}{|c|c|c|c|}
\colrule
Solution & Energy limit& $\chi^2$/N$_{\rm Data}$ & N$_{\rm Data}$ \\
         & (MeV)       &               &      \\
\colrule
GZ12     & 2700        &  1.95         & 26179 \\
GB12     & 2700        &  2.09         & 26179 \\
\colrule
SN11     & 2700        &  2.08         & 25553 \\
SP09     & 2700        &  2.05         & 24912 \\
SM02     & 2000        &  2.01         & 17571 \\
SM95     & 2000        &  2.37         & 13415 \\
\colrule
\end{tabular}
\end{table}
In Table~\ref{tab:tbl1}, we compare the new GB12 and GZ12 solutions 
with four previous SAID fits (SN11~\cite{sn11}, SP09~\cite{du1}, 
SM02~\cite{SAID02}, and SM95~\cite{sm95}). The overall $\chi^2$ has 
remained stable against the growing database, which has increased 
by a factor of 2 since 1995 (most of this increase coming from 
data from photon-tagging facilities).

In fitting the data, the stated experimental systematic 
uncertainties have been used as an overall normalization adjustment 
factor for the angular distributions~\cite{SAID02,sn11}. Presently, 
the pion photoproduction database below E$_\gamma$ = 2.7~GeV 
consists of 26179 data points that have been fit in the GB12 (GZ12) 
solution with $\chi^2$ = 54832 (50998).  The contribution to the 
total $\chi^2$ in the GB12 (GZ12) analyses of the $626$ new CLAS 
$\gnp$ data points (e.g., those data points up to E$_\gamma$ = 
2.7~GeV) is 1580 (1190).  This should be compared to a starting 
$\chi^2$ = 45636 for the new CLAS data using predictions from
our previous SN11 solution. 

The solution, GB12, uses the same fitting form as our recent SN11 
solution~\cite{sn11}, which incorporated the neutron-target $\Sigma$ 
data from GRAAL~\cite{graal1,graal2}. This fit form was motivated by 
a multi-channel K-matrix approach, with an added phenomenological 
term proportional to the $\pi N$ reaction cross section.  A second 
fit, GZ12, used instead the recently proposed form~\cite{mc12} based 
on a unified Chew-Mandelstam parameterization of the GW DAC fits to 
both $\pi N$ elastic scattering and photoproduction. This form 
explicitly includes contributions from channels $\pi N$, $\pi\Delta$, 
$\rho N$, and $\eta N$, as determined in the SAID elastic $\pi N$ 
scattering analysis. 

\begin{figure*}[th]
\centerline{
\includegraphics[height=0.42\textwidth, angle=90]{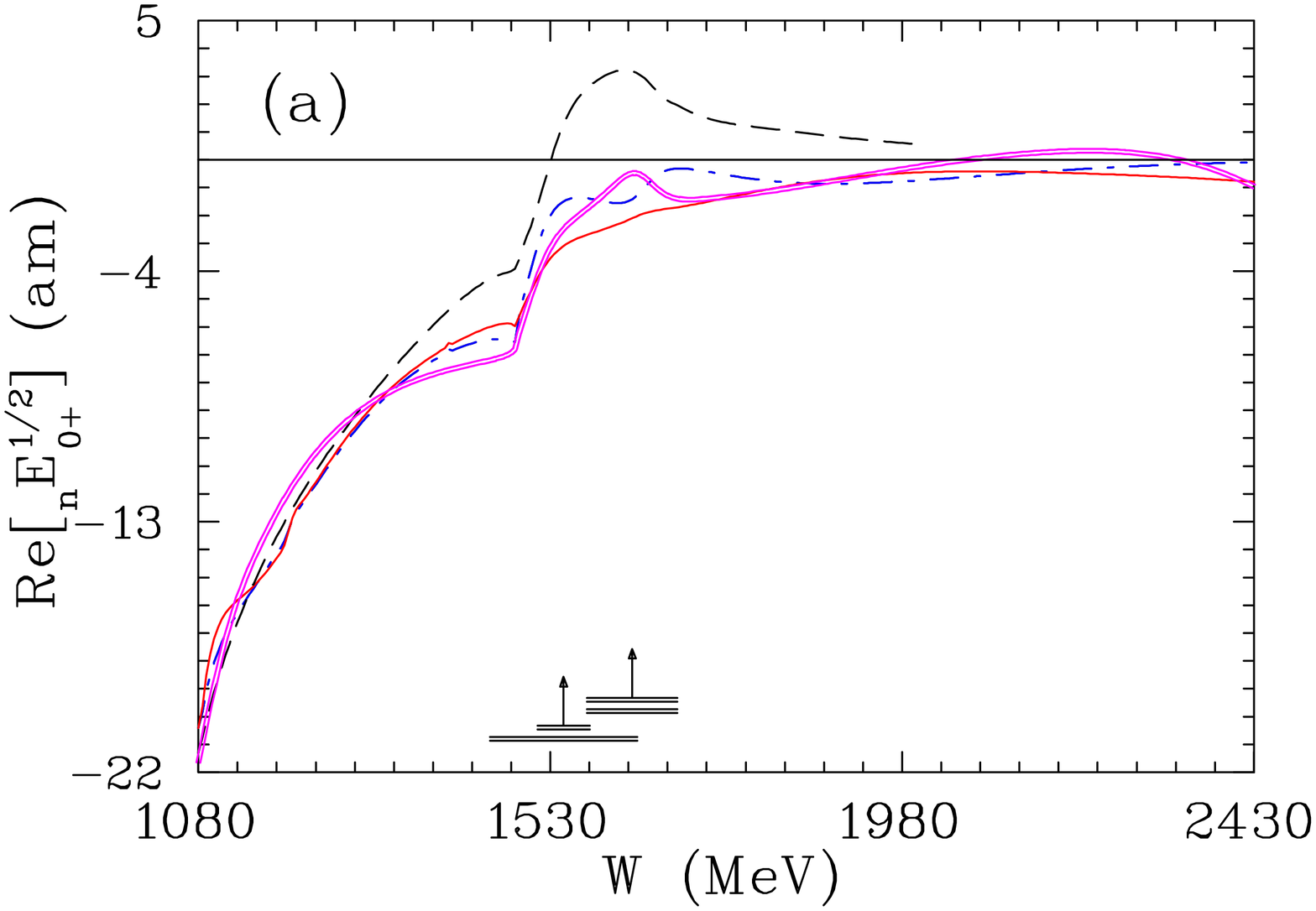}\hfill
\includegraphics[height=0.42\textwidth, angle=90]{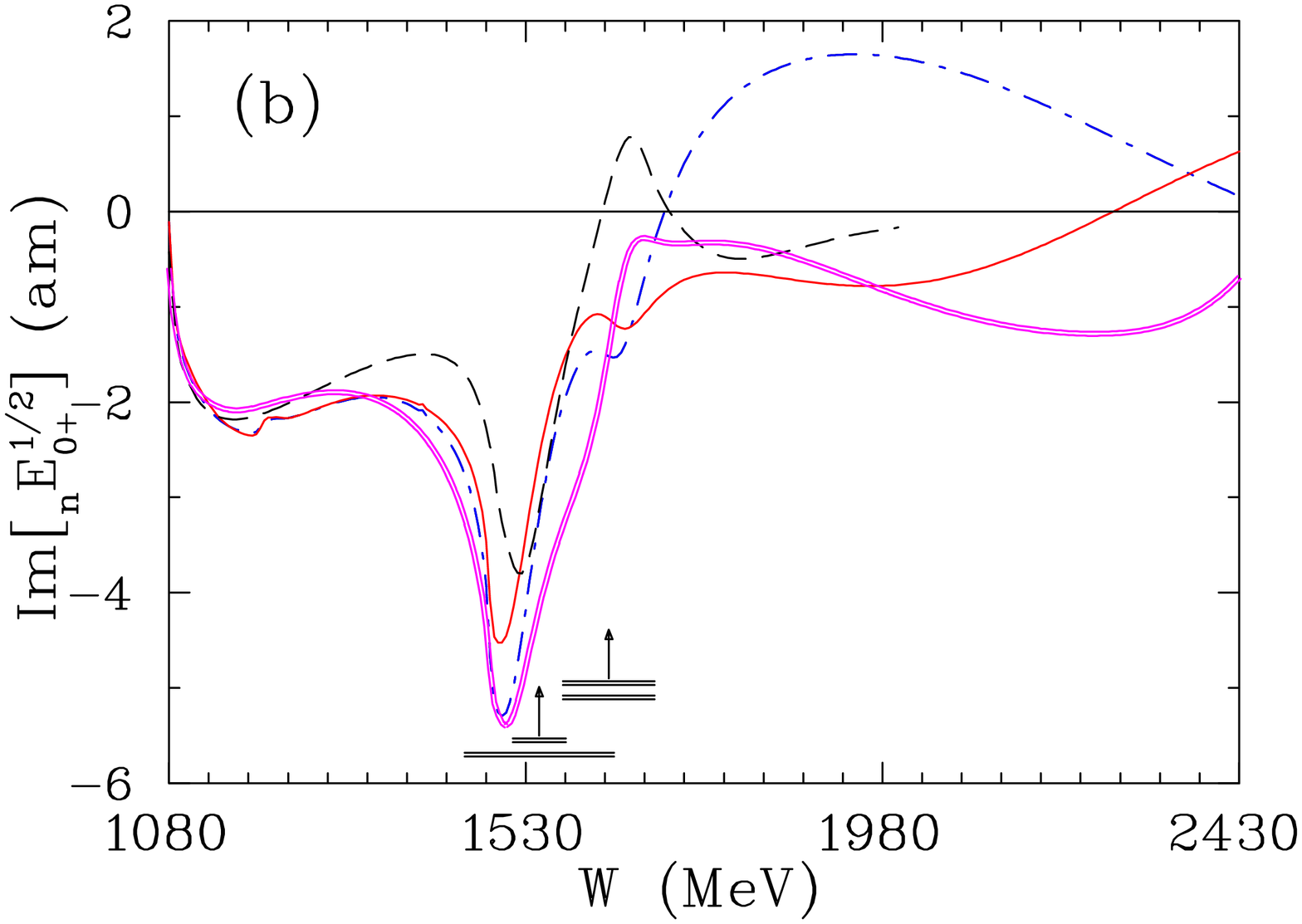}}
\centerline{
\includegraphics[height=0.42\textwidth, angle=90]{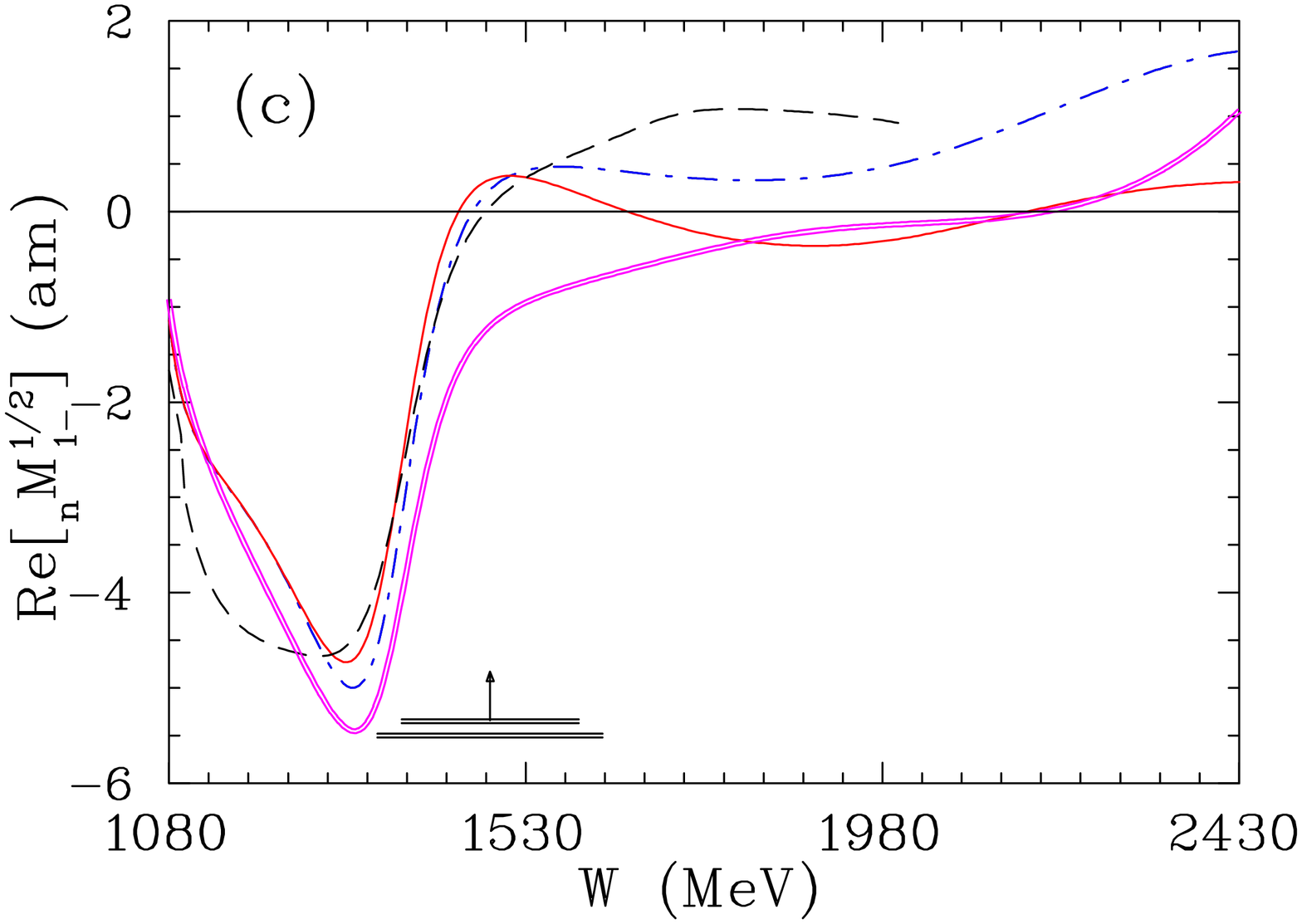}\hfill
\includegraphics[height=0.42\textwidth, angle=90]{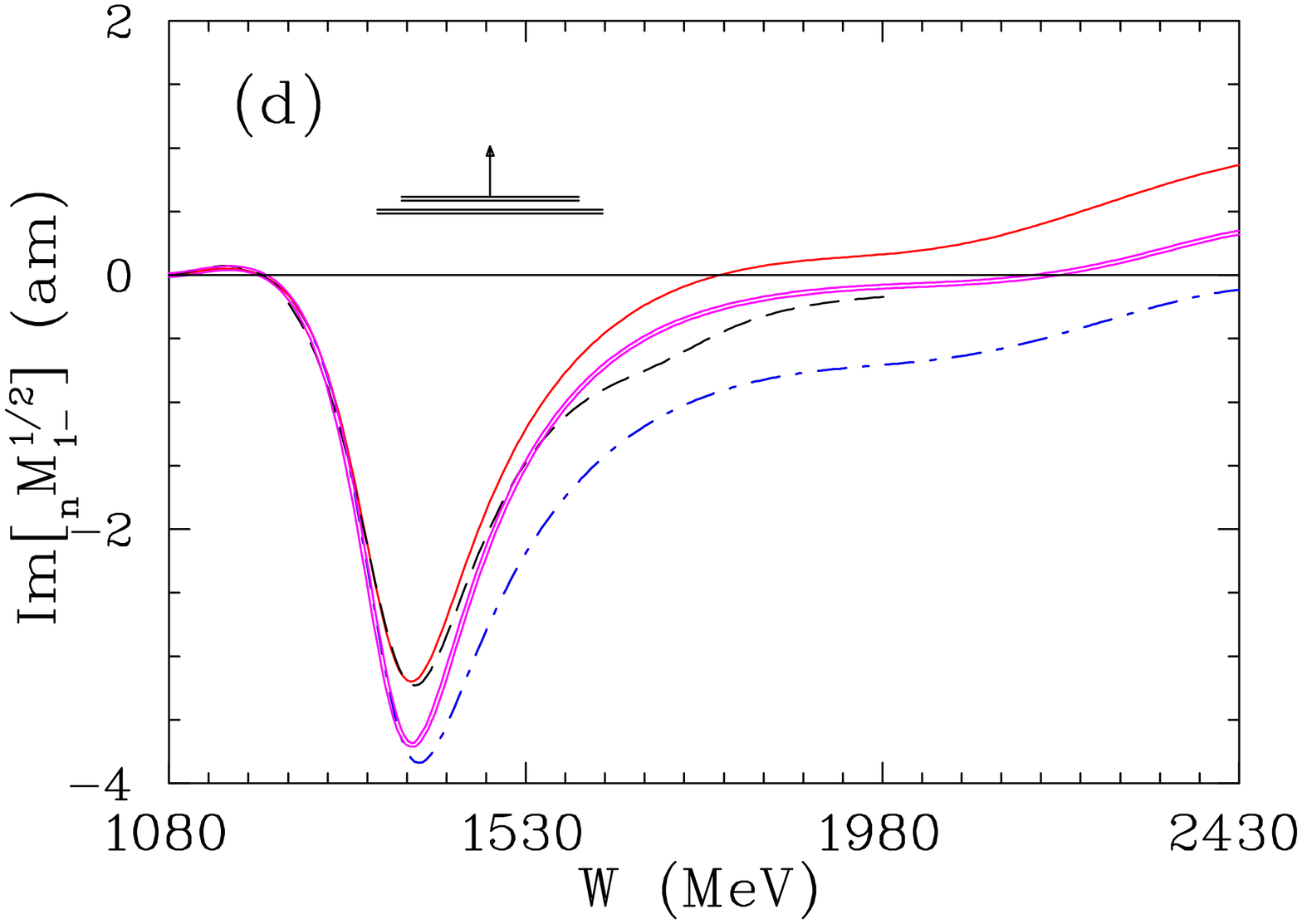}}
\centerline{
\includegraphics[height=0.42\textwidth, angle=90]{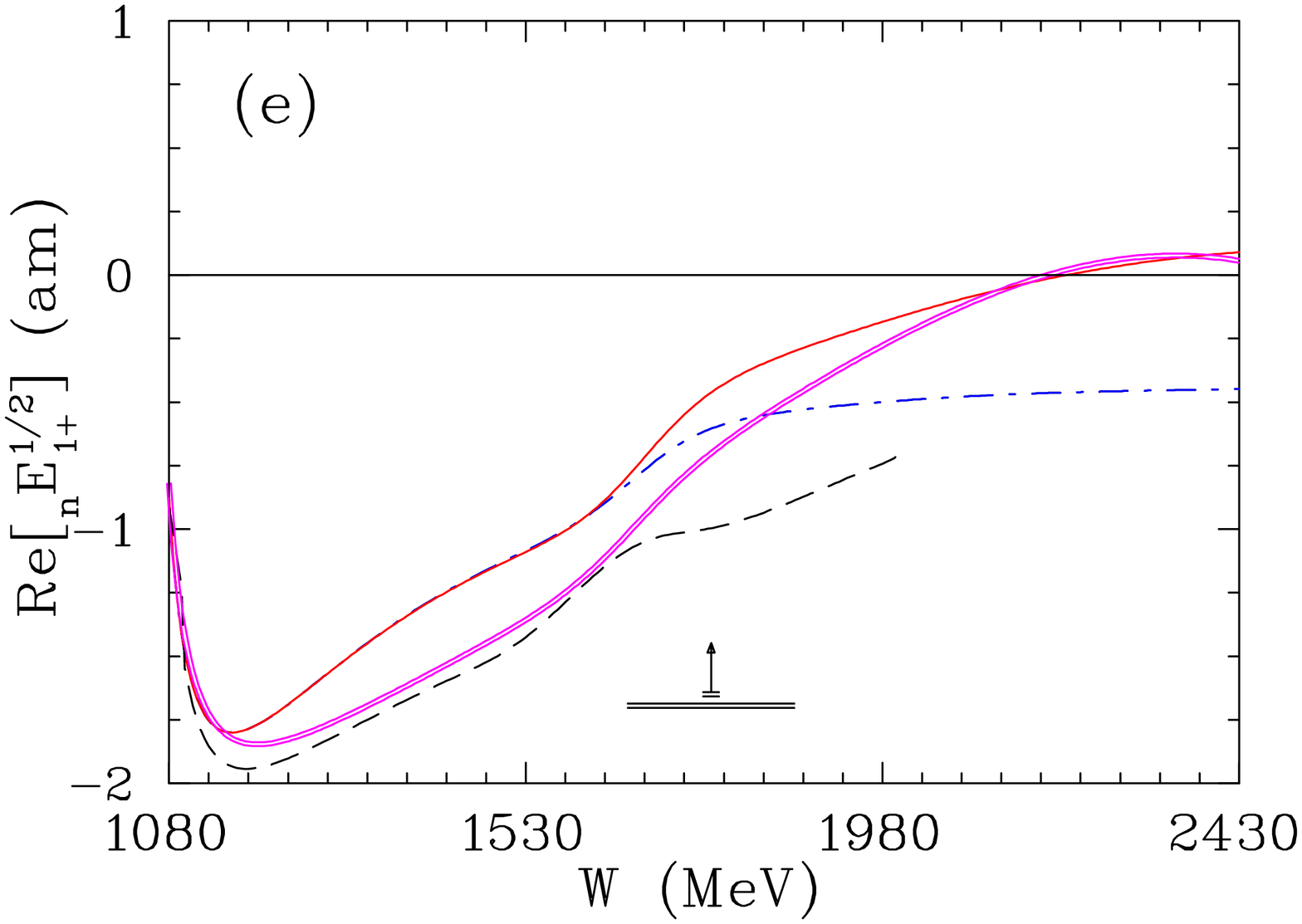}\hfill
\includegraphics[height=0.42\textwidth, angle=90]{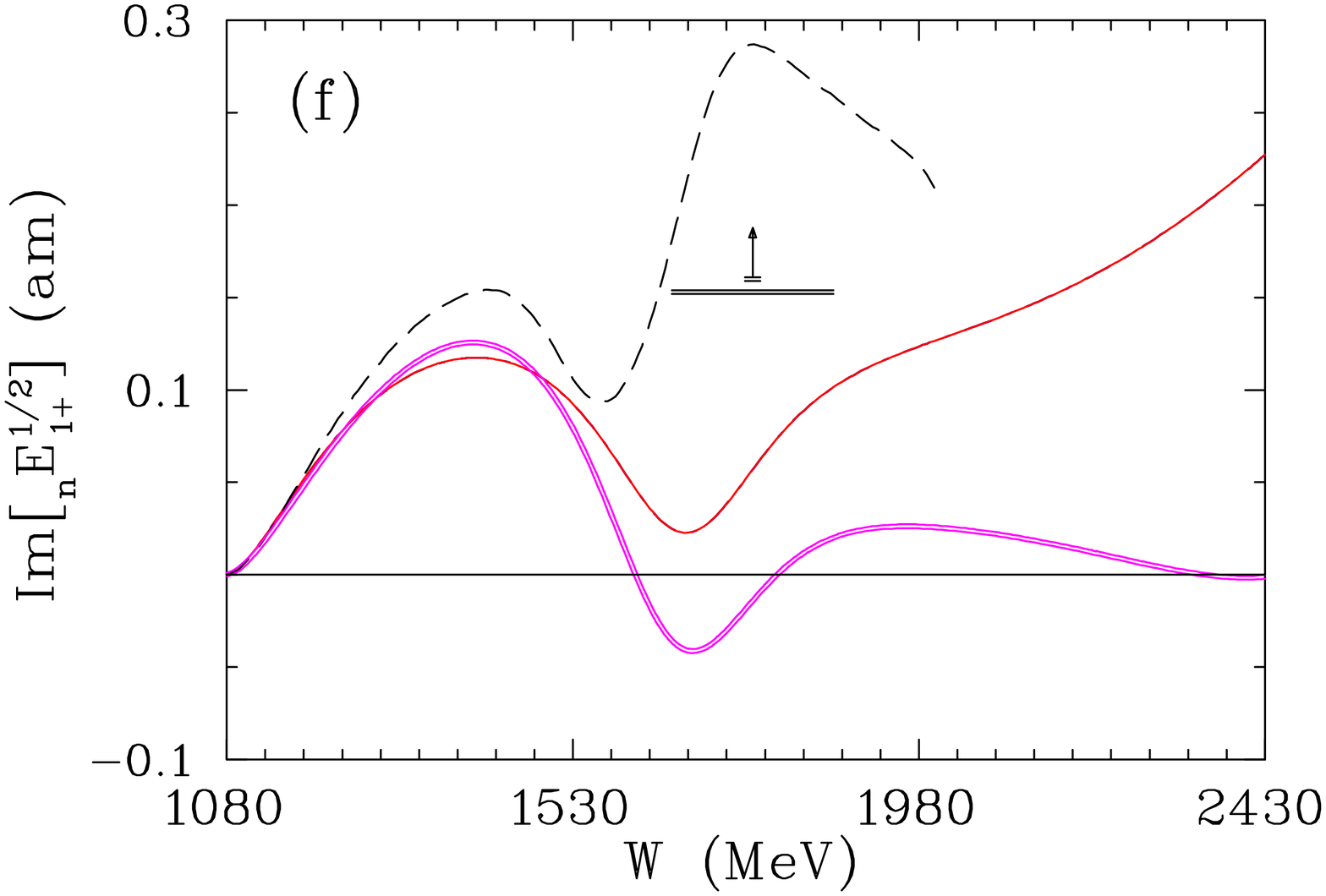}}
\centerline{
\includegraphics[height=0.42\textwidth, angle=90]{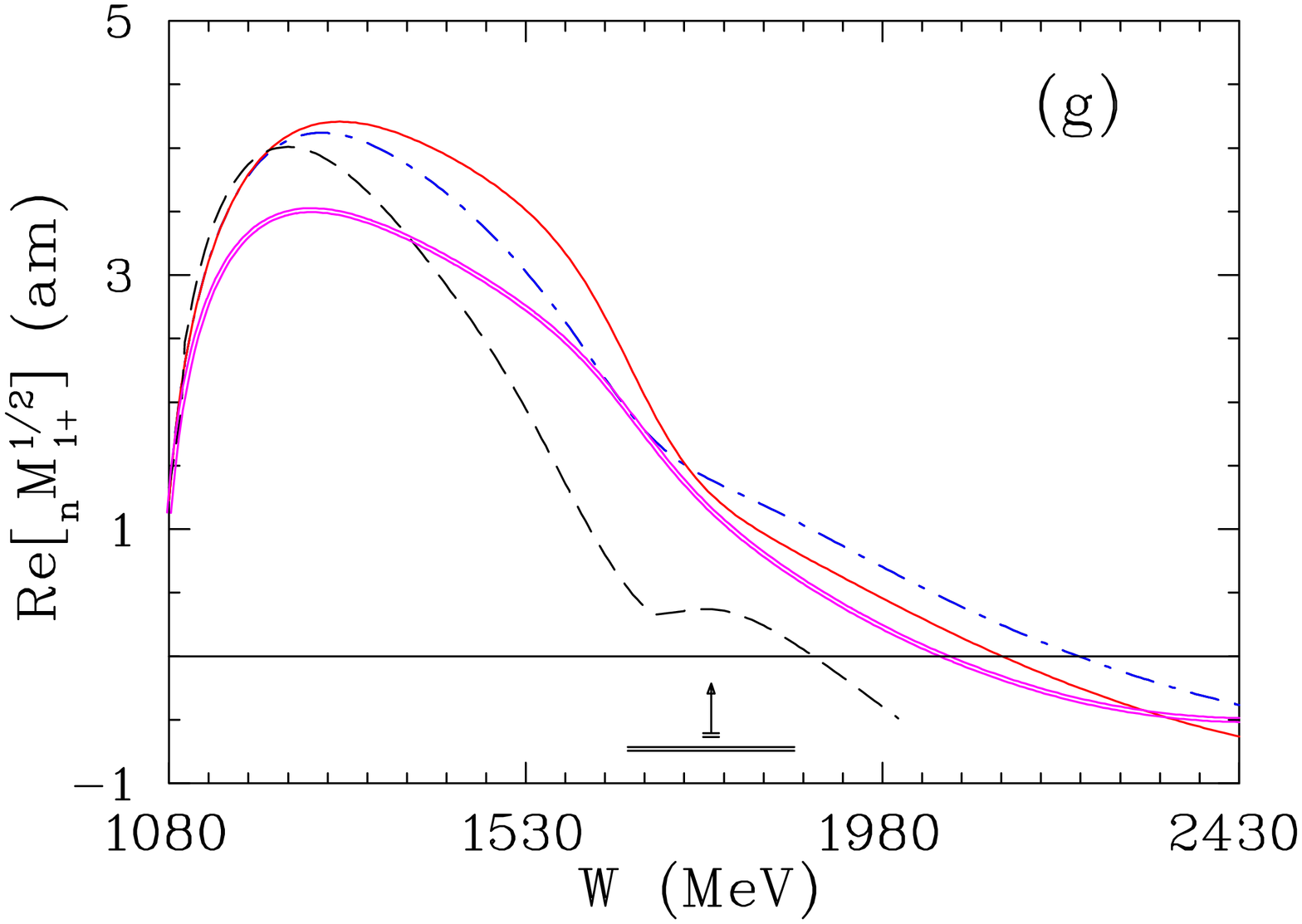}\hfill
\includegraphics[height=0.42\textwidth, angle=90]{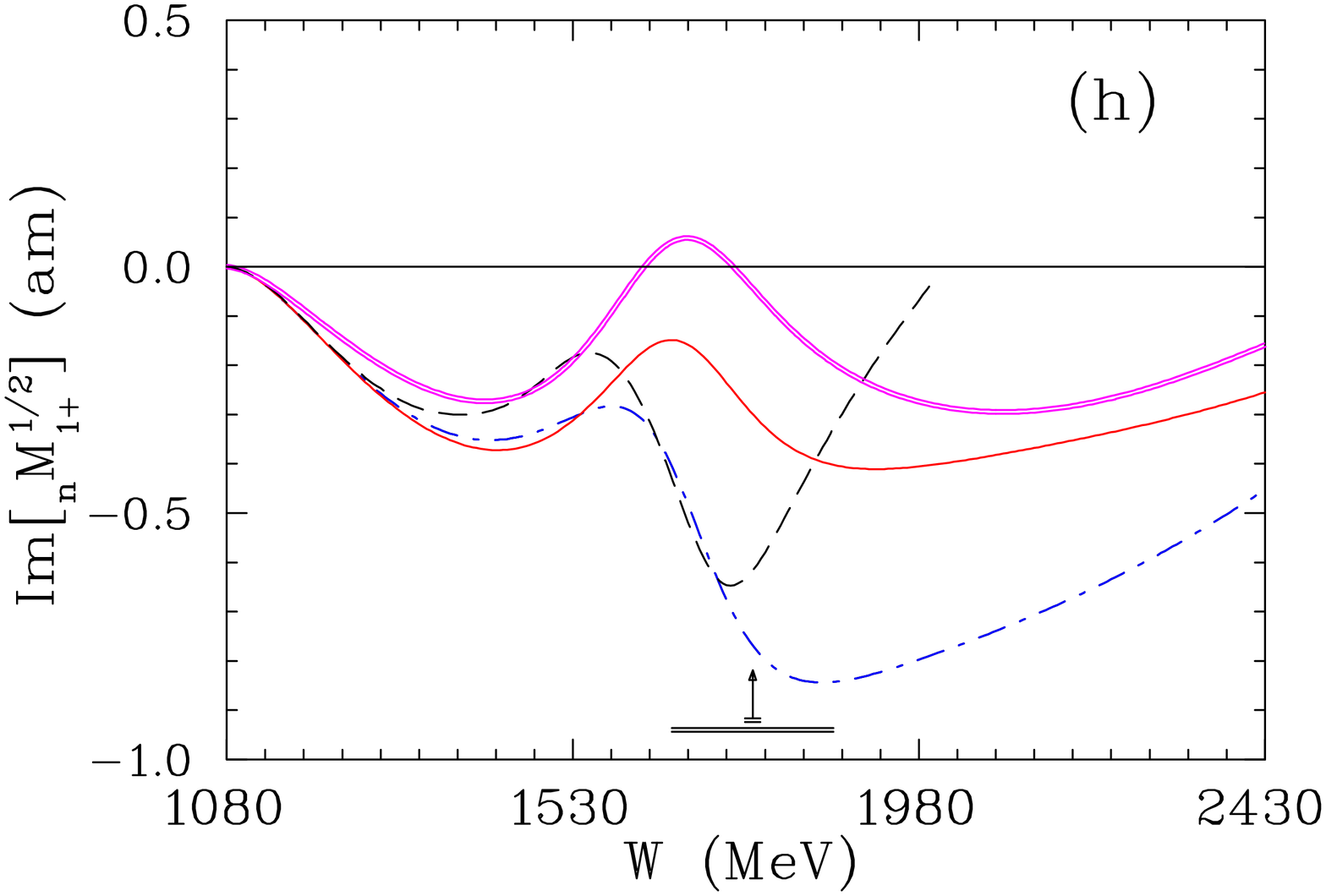}}
\caption{(Color online) Neutron multipole I=1/2 amplitudes from
        threshold to $W$ = 2.43~GeV ($E_{\gamma}$ = 2.7~GeV).
	Solid (dash-dotted) lines correspond to the GB12 
	(SN11~\protect\cite{sn11}) solution.  Thick solid (dashed) 
	lines give GZ12 solution (MAID07~\protect\cite{maid},
        which terminates at $W$ = 2~GeV). Vertical arrows indicate 
	resonance energies, $W_R$, and horizontal bars show full 
	($\Gamma$) and partial ($\Gamma_{\pi N}$) widths 
	associated with the SAID $\pi N$ solution SP06
	\protect\cite{piN_PWA}. \label{fig:g5}}
\end{figure*}

\begin{figure*}[th]
\centerline{
\includegraphics[height=0.42\textwidth, angle=90]{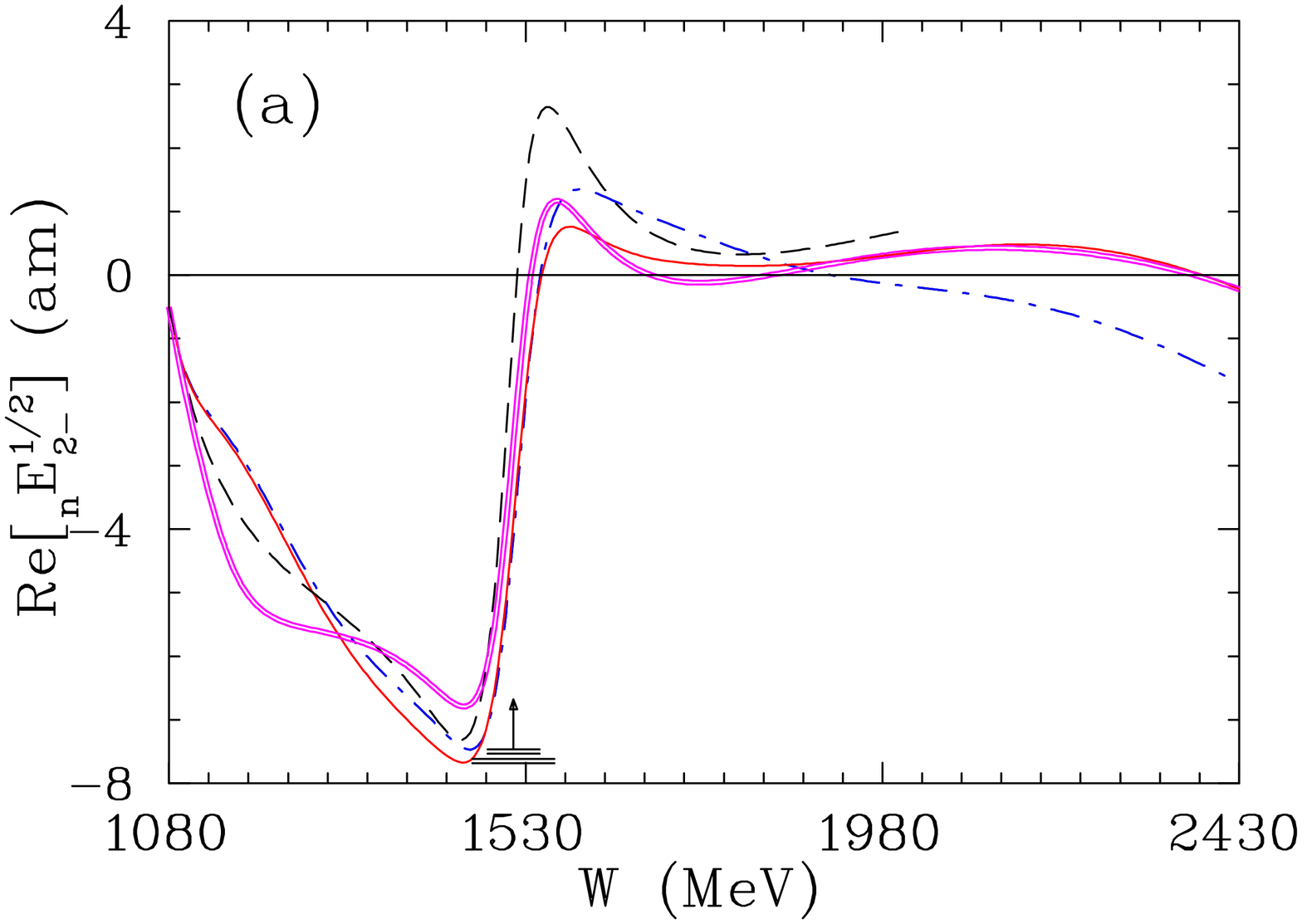}\hfill
\includegraphics[height=0.42\textwidth, angle=90]{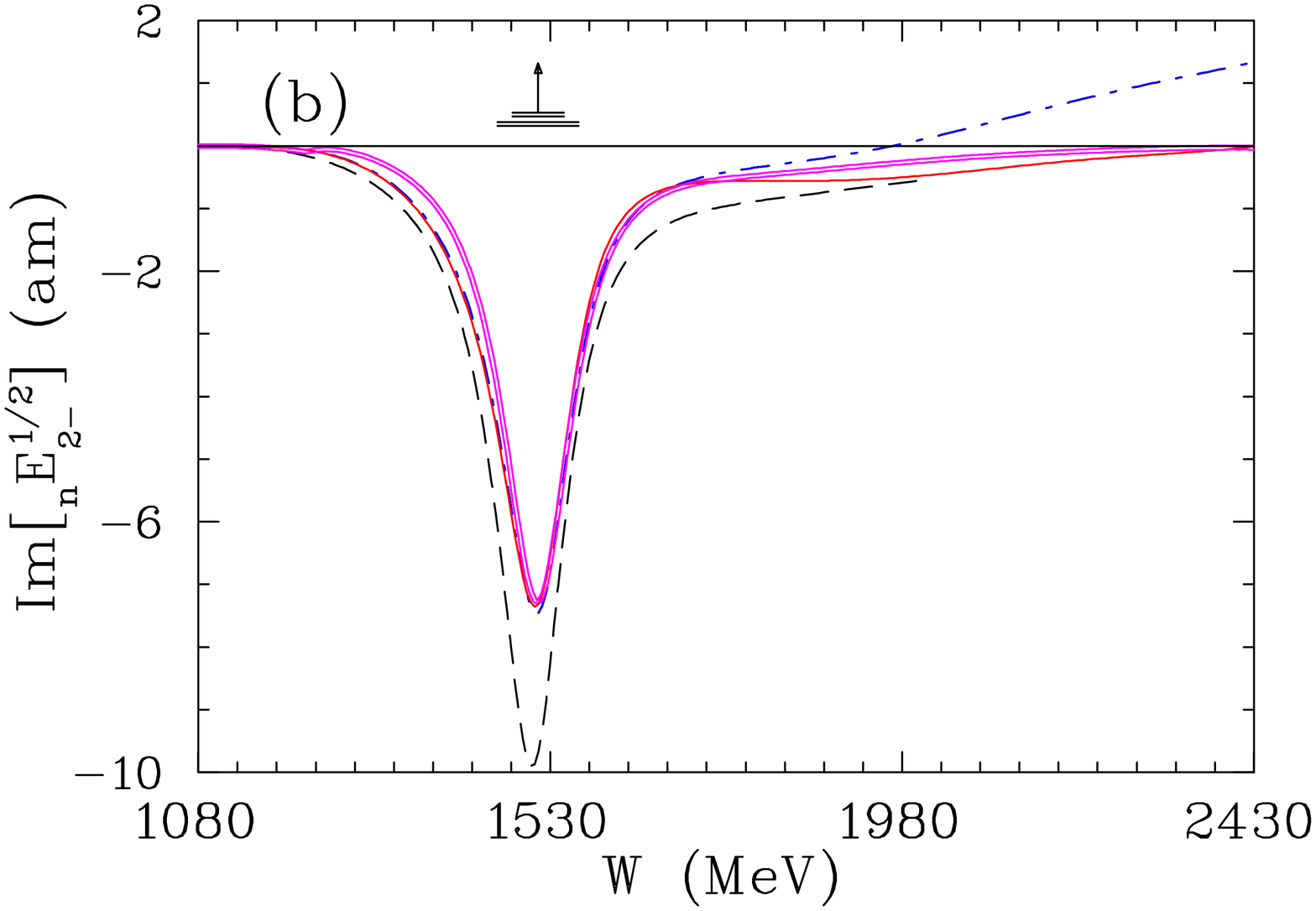}}
\centerline{
\includegraphics[height=0.42\textwidth, angle=90]{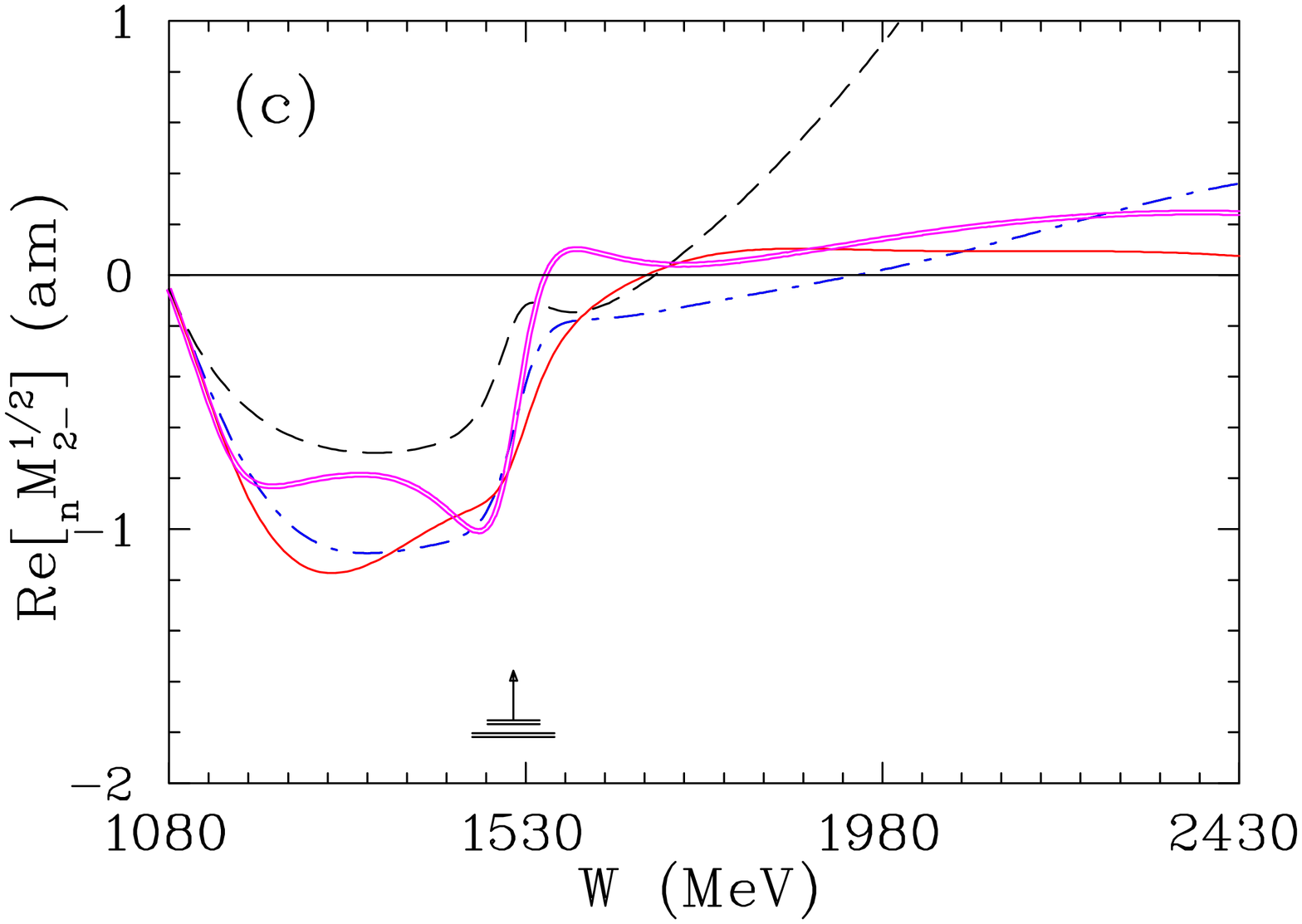}\hfill
\includegraphics[height=0.42\textwidth, angle=90]{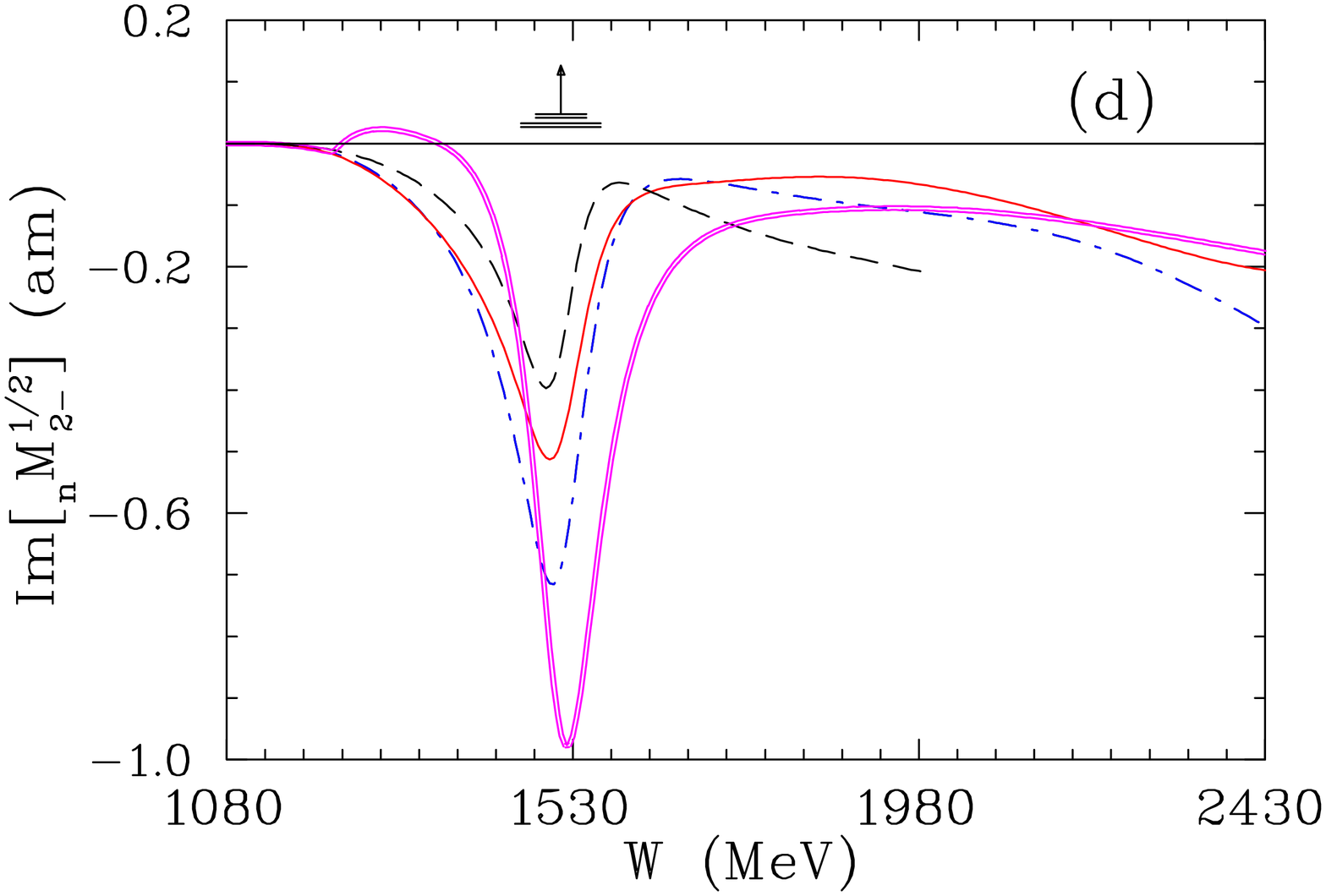}}
\centerline{
\includegraphics[height=0.42\textwidth, angle=90]{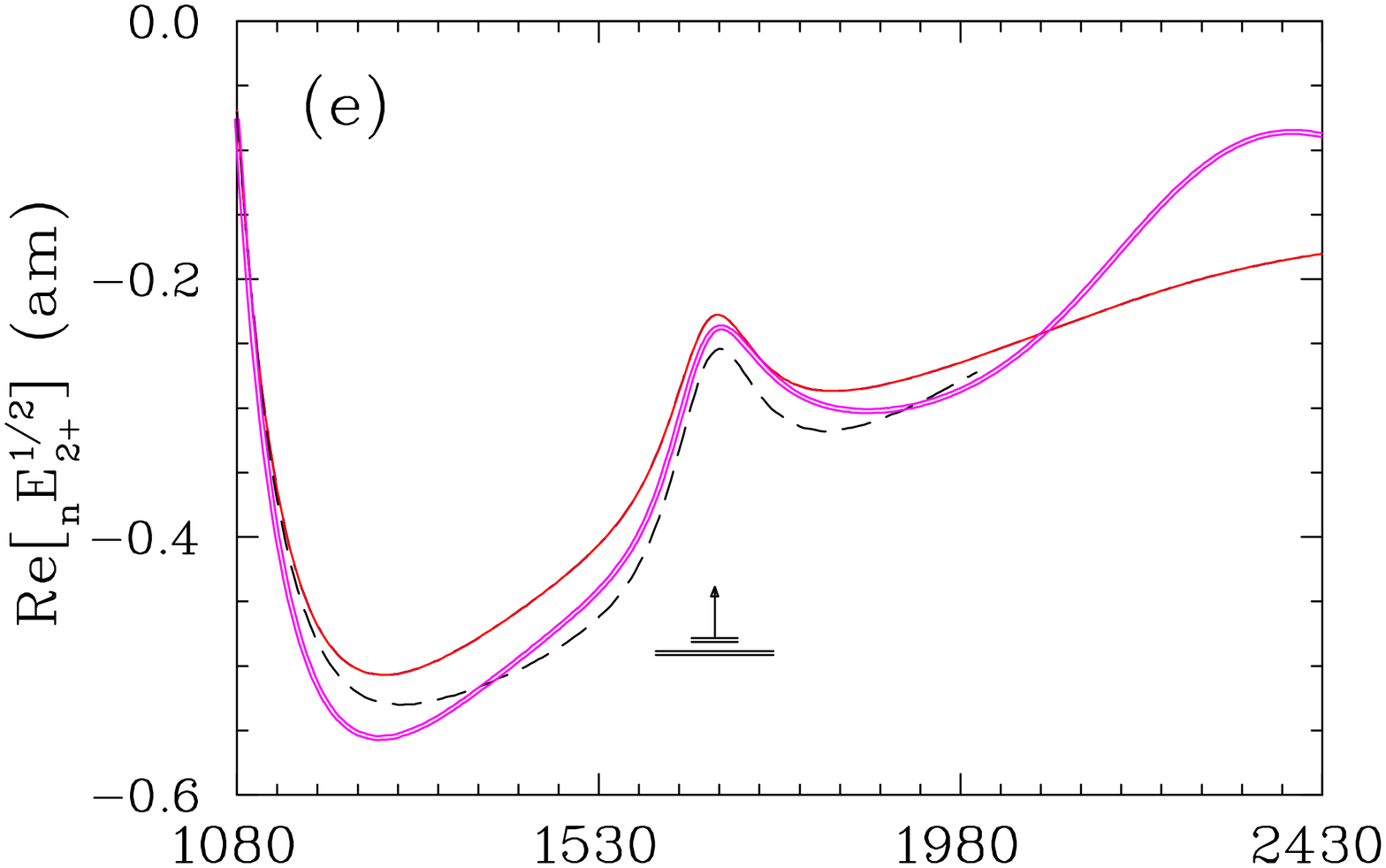}\hfill
\includegraphics[height=0.42\textwidth, angle=90]{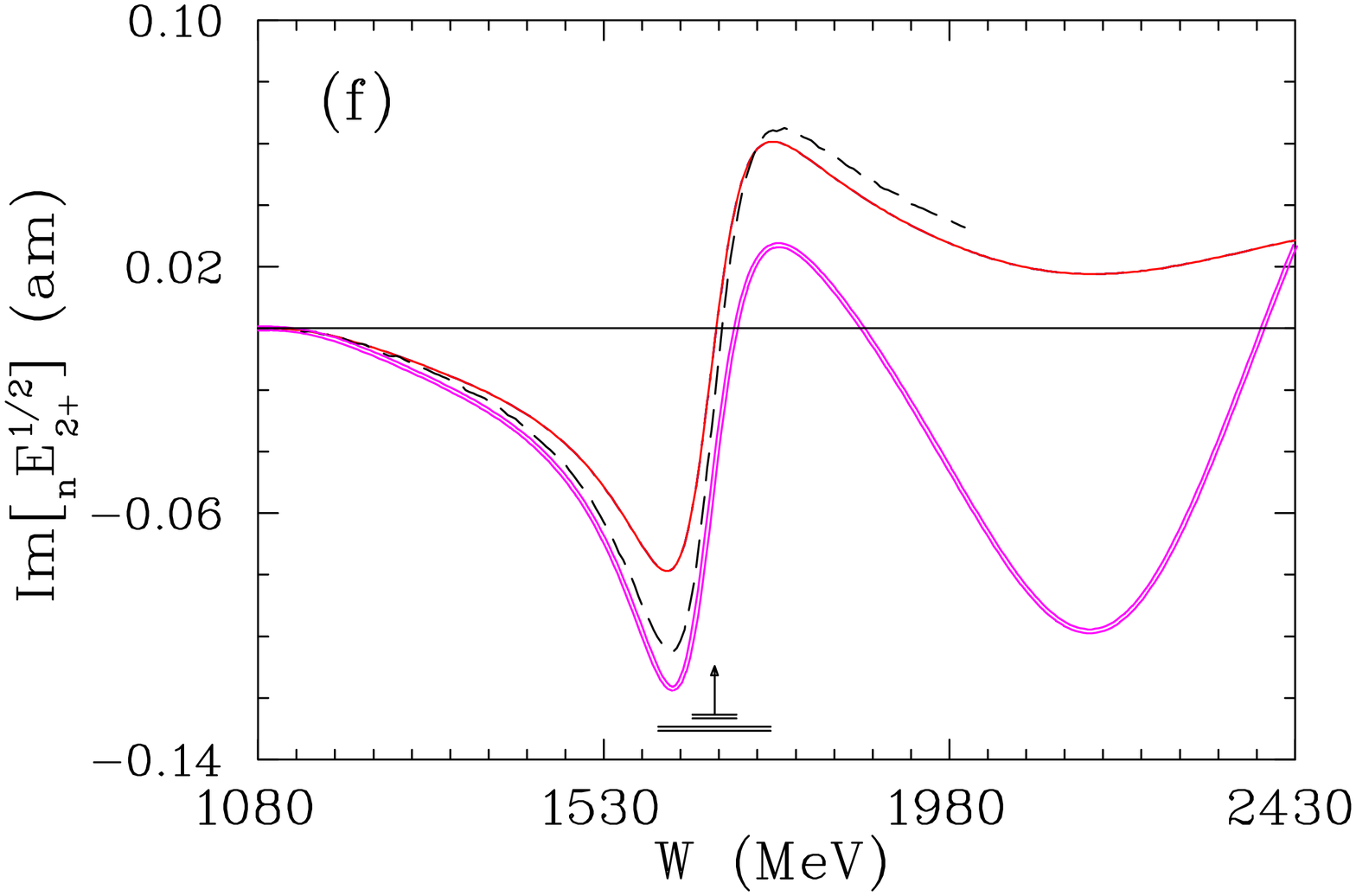}}
\centerline{
\includegraphics[height=0.42\textwidth, angle=90]{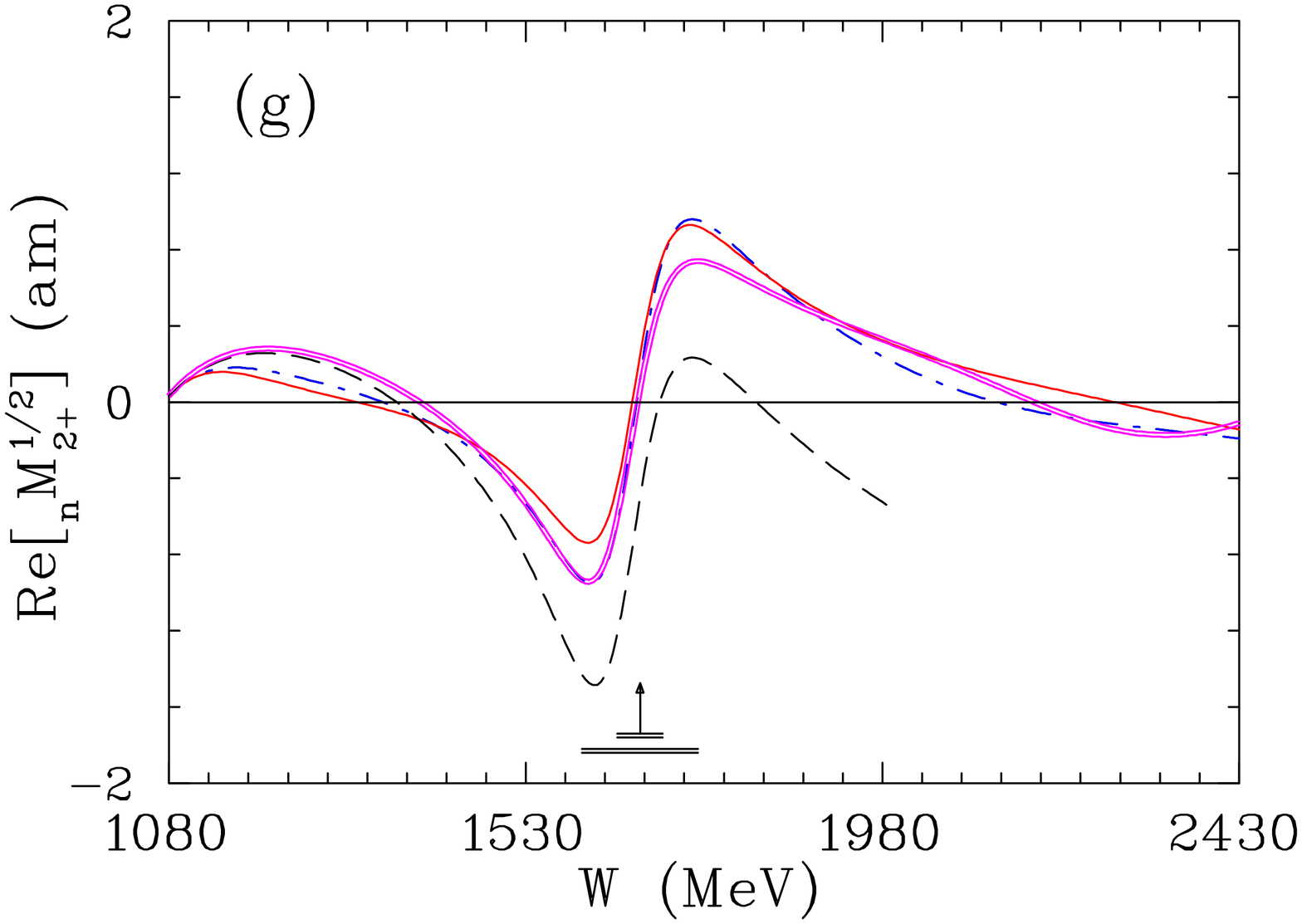}\hfill
\includegraphics[height=0.42\textwidth, angle=90]{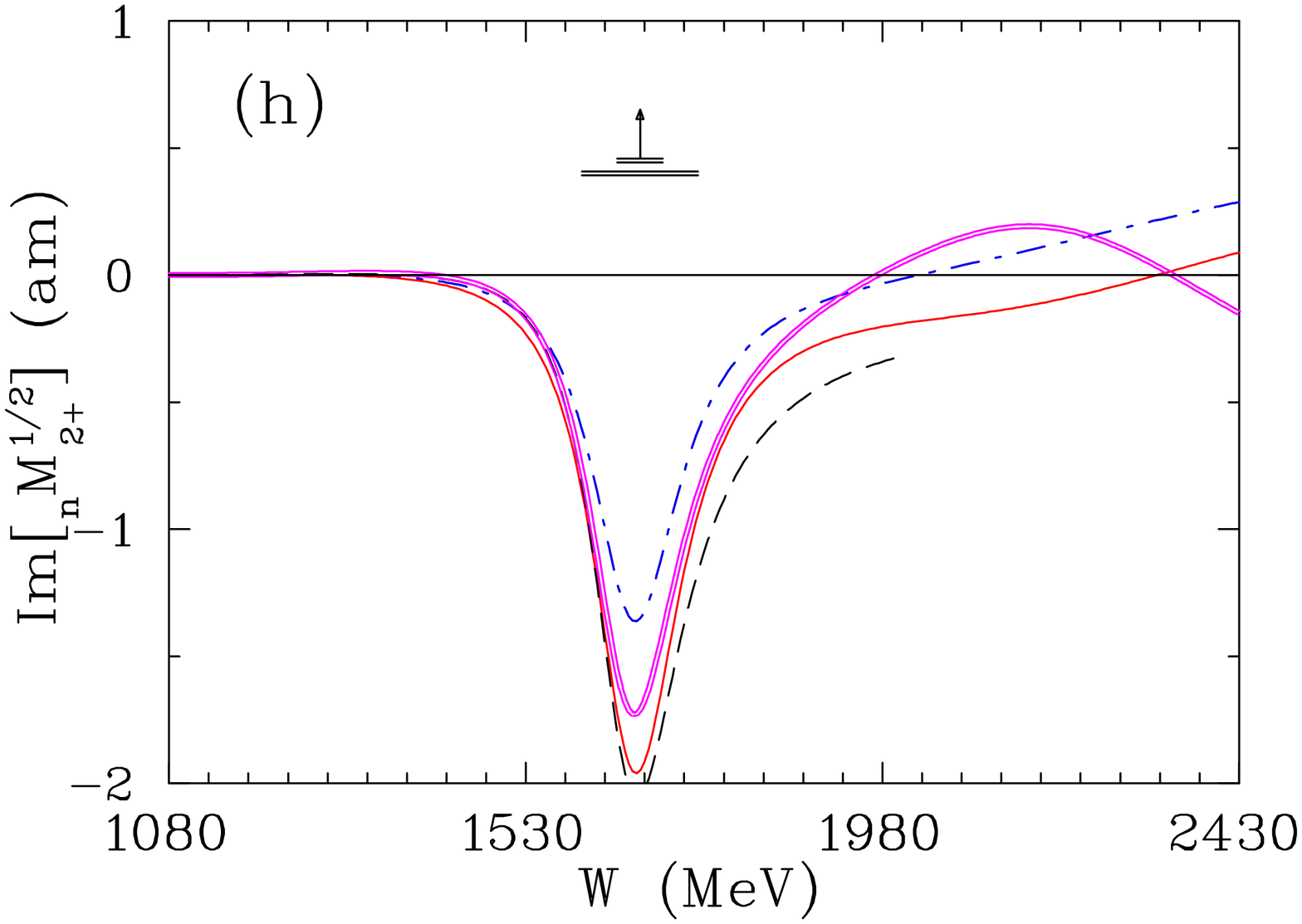}}
\caption{(Color online) Notation of the multipoles is the
        same as in Fig.~\protect\ref{fig:g5}. \label{fig:g6}}
\end{figure*}

\begin{figure*}[th]
\centerline{
\includegraphics[height=0.42\textwidth, angle=90]{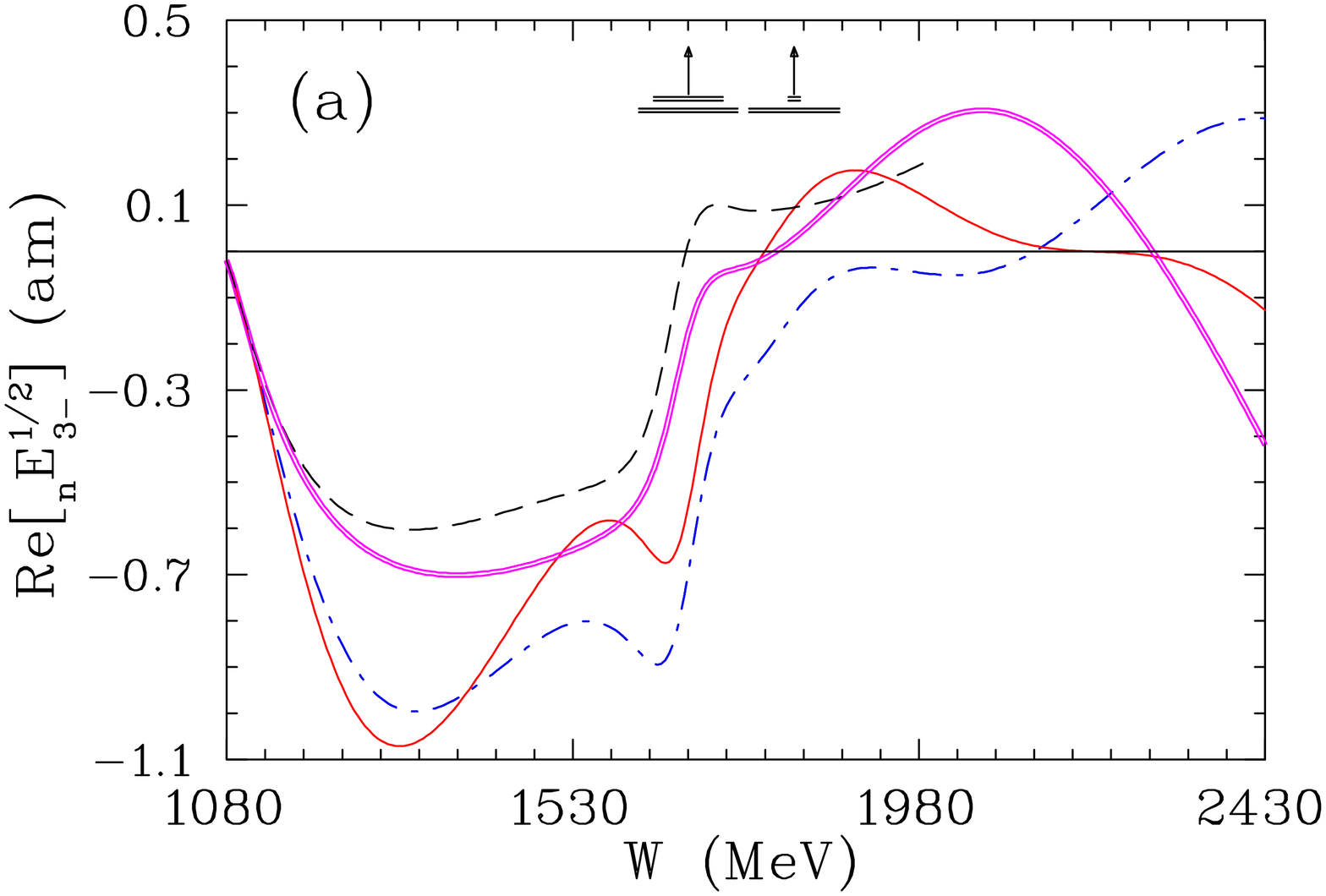}\hfill
\includegraphics[height=0.42\textwidth, angle=90]{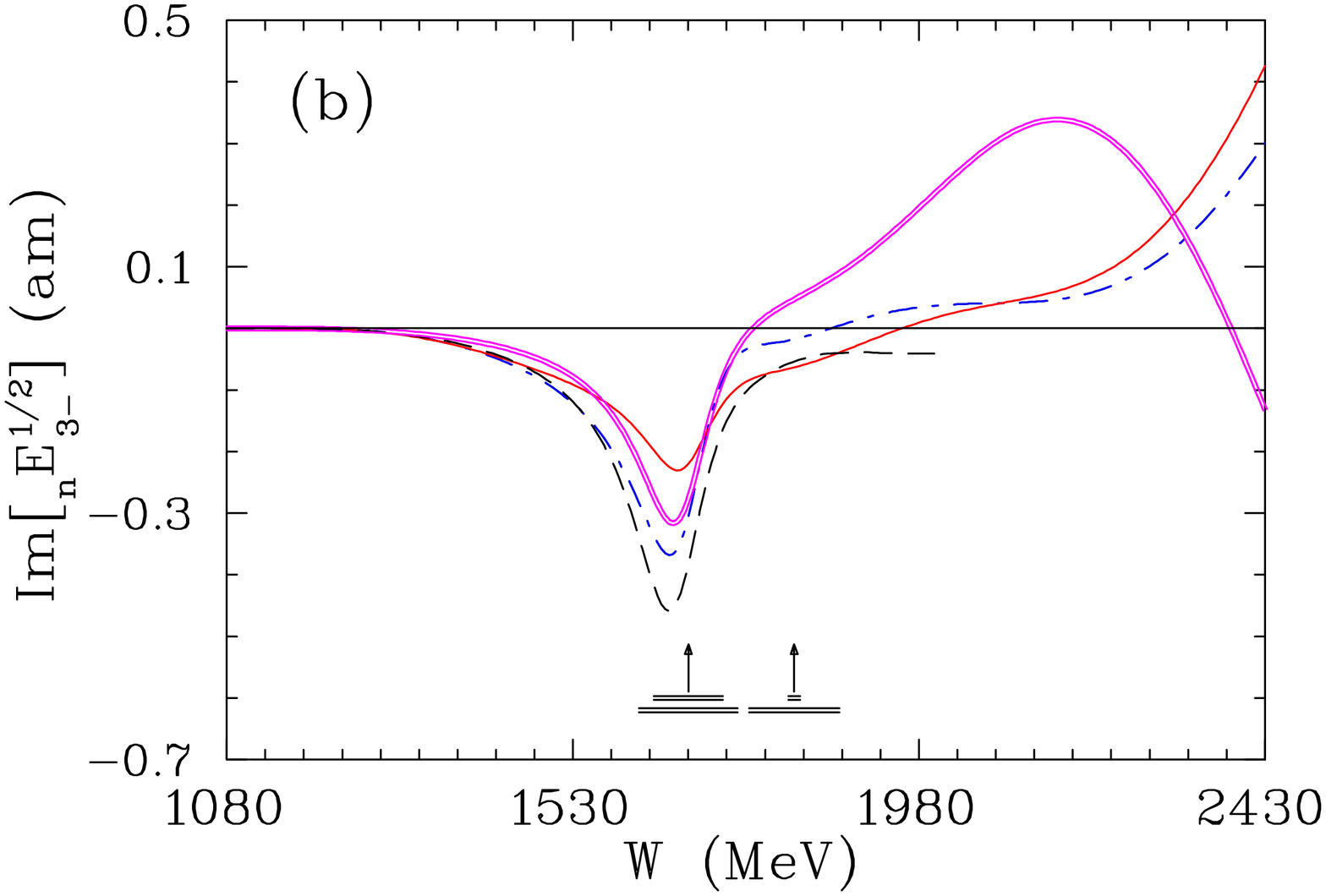}}
\centerline{
\includegraphics[height=0.42\textwidth, angle=90]{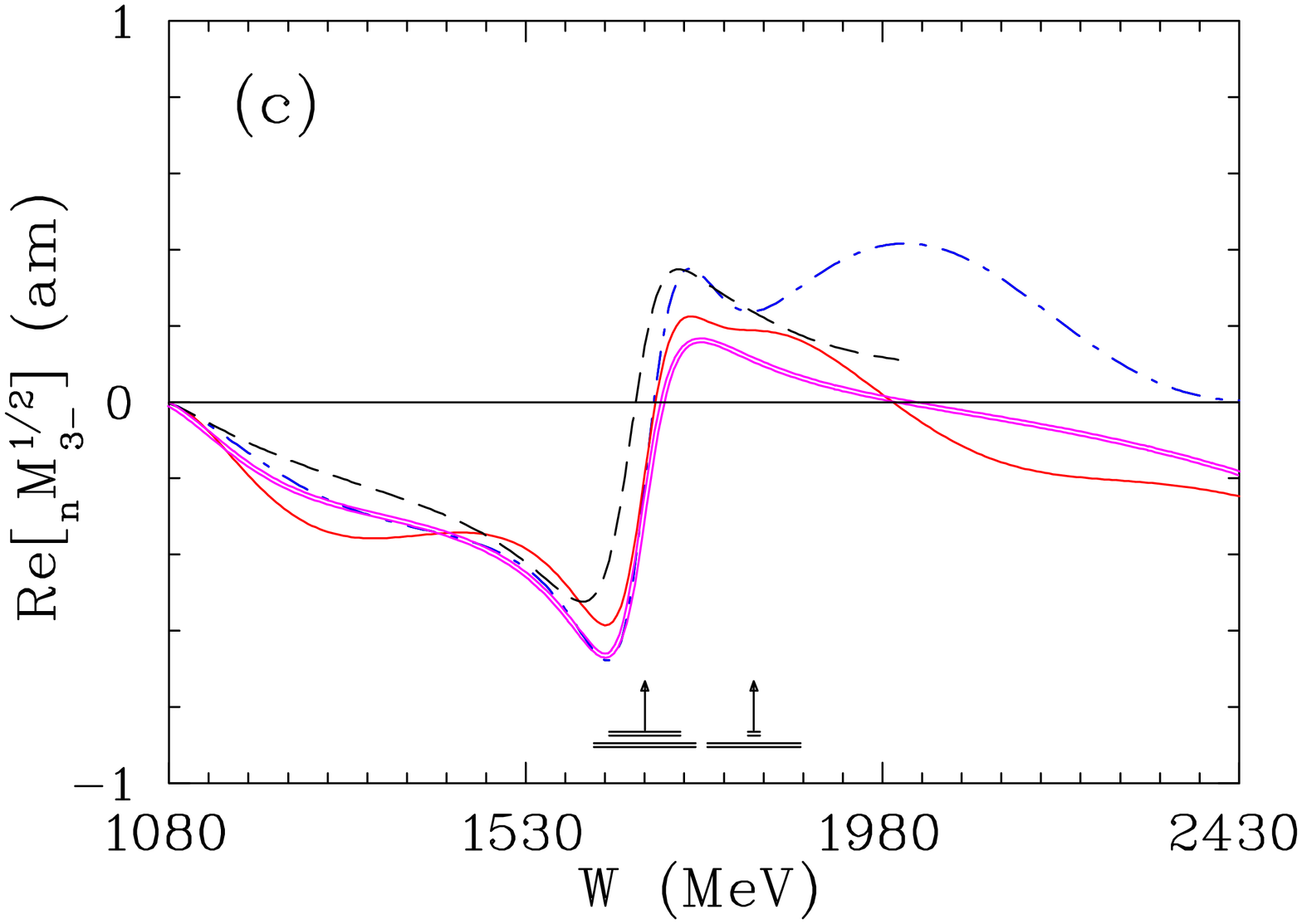}\hfill
\includegraphics[height=0.42\textwidth, angle=90]{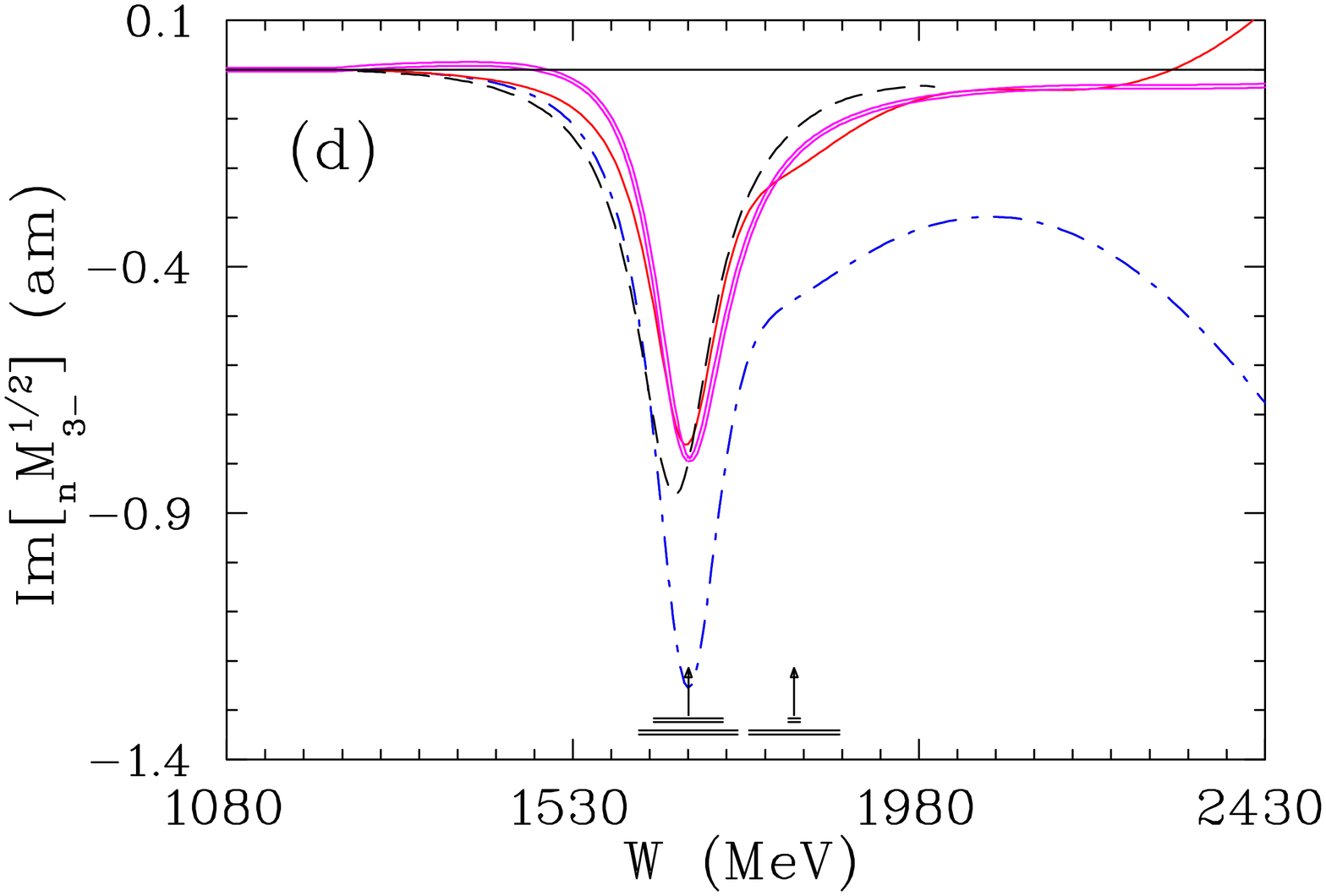}}
\caption{(Color online) Notation of the multipoles is the
        same as in Fig.~\protect\ref{fig:g5}.\label{fig:g7}}
\end{figure*}

Resonance couplings, extracted as in Ref.~\cite{sn11}, are listed in 
Table~\ref{tab:tbl2} and compared to the previous SN11 determinations 
and the Particle Data Group (PDG) averages~\cite{PDG}. Couplings for 
the $N(1440)1/2^+$, $N(1520)3/2^-$, and $N(1675)5/2^-$ are reasonably 
close to the SN11 estimates. The value of $A_{1/2}$ found for the 
$N(1535)1/2^-$, using the GB12 fit, is very close to the SN11 
determination. Using the GZ12 fit, however, the result is somewhat 
larger in magnitude ($-85\pm 15$). A similar feature was found for 
the proton couplings, using this form, in Ref.~\cite{mc12}.  Using 
this alternate form, a determination of the $N(1650)1/2^-$ $A_{1/2}$ 
was difficult and resulted in a value, lower in magnitude by about 
50\%, compared to the value from GB12 listed in Table~\ref{tab:tbl2}. 
For this reason, we consider the uncertainty associated with this 
state to be a lower limit only. No value was quoted for the 
$N(1720)3/2^+$ state. As can be seen in Figs.~\ref{fig:g5} -- 
\ref{fig:g7}, the two different fit forms GB12 and GZ12, though 
similar in shape, have opposite signs for the imaginary parts of 
corresponding multipoles ($_n E^{1/2}_{1+}$ and $_n M^{1/2}_{1+}$) 
in the neighborhood of the resonance position, and even the sign 
can not be determined.  This is in line with the PDG estimates, 
which also fail to give signs for the couplings to this state.


\begin{table}[th]
\caption{Resonance parameters for N$^\ast$ states
         from the SAID fit to the $\pi N$ data ~\protect\cite{piN_PWA}
         (second column) and neutron helicity amplitudes $A_{1/2}$ and
         $A_{3/2}$ (in [(GeV)$^{-1/2}\times 10^{-3}$] units) from the 
	 GB12 solution (first row), previous SN11~\protect\cite{sn11}
         solution (second row), and average values from the
         PDG10~\protect\cite{PDG} (third row). $^{\dagger}$See text.
         \label{tab:tbl2}}
\vspace{2mm}
\begin{tabular}{|c|c|c|c|}
\colrule
Resonance       & $\pi N$ SAID               &   $A_{1/2}$  & $A_{3/2}$    \\
\colrule
$N(1535)1/2^-$  & $W_{R}$=1547~MeV           &  $-$58$\pm$6$^\dagger$  &   \\
                & $\Gamma$=188~MeV           &  $-$60$\pm$3 & \\
                & $\Gamma _{\pi}/\Gamma$=0.36&  $-$46$\pm$27& \\
\colrule
$N(1650)1/2^-$  & $W_{R}$=1635~MeV           &  $-$40$\pm$10$^\dagger$  &  \\
                & $\Gamma$=115~MeV           &  $-$26$\pm$8 & \\
                & $\Gamma _{\pi}/\Gamma$=1.00& $-$15$\pm$21 & \\
\colrule
$N(1440)1/2^+$  & $W_{R}$=1485~MeV           &   48$\pm$4   & \\
                & $\Gamma$=284~MeV           &   45$\pm$15  & \\
                & $\Gamma _{\pi}/\Gamma$=0.79&  40$\pm$10   & \\
\colrule
$N(1520)3/2^-$  & $W_{R}$=1515~MeV           & $-$46$\pm$6  & $-$115$\pm$5 \\
                & $\Gamma$=104~MeV           & $-$47$\pm$2  & $-$125$\pm$2 \\
                & $\Gamma _{\pi}/\Gamma$=0.63&$-$59$\pm$9   & $-$139$\pm$11\\
\colrule
$N(1675)5/2^-$  & $W_{R}$=1674~MeV           & $-$58$\pm$2  & $-$80$\pm$5  \\
                & $\Gamma$=147~MeV           & $-$42$\pm$2  & $-$60$\pm$2  \\
                & $\Gamma _{\pi}/\Gamma$=0.39& $-$43$\pm$12 & $-$58$\pm$13 \\
\colrule
$N(1680)5/2^+$  & $W_{R}$=1680~MeV           &26$\pm$4      & $-$29$\pm$2  \\
                & $\Gamma$=128~MeV           &50$\pm$4      & $-$47$\pm$2  \\
                & $\Gamma _{\pi}/\Gamma$=0.70&29$\pm$10     & $-$33$\pm$9  \\
\colrule
\end{tabular}
\end{table}

\section{Summary and Conclusion}
\label{sec:conc}

A comprehensive set of differential \crss\ at 26 energies for 
negative-pion photoproduction on the neutron, via the reaction $\gdpp$, 
have been determined with a JLab tagged-photon beam for incident 
photon energies from 1.05 to 3.5~GeV. To accomplish a state-of-the-art 
analysis, we included new FSI corrections using a diagrammatic 
technique, taking into account a kinematical cut with momenta less 
(more) than 200~MeV/$c$ for slow (fast) outgoing protons.

The updated PWAs examined mainly the effect of new CLAS neutron-target 
data on the SAID multipoles and resonance parameters. These new data 
have been included in a SAID multipole analysis, resulting in new SAID 
solutions, GB12 and GZ12. A major accomplishment of this CLAS 
experiment is a substantial improvement in the $\pi^-$-photoproduction 
database, adding $855$ new differential \crss\, quadrupling the 
world database for $\gnp$ above 1~GeV.  Comparison to earlier SAID 
fits, and a lower-energy fit from the Mainz group, 
shows that the new solutions are 
much more satisfactory at higher energies.

On the experimental side, further improvements in the PWAs await more
data, specifically in the region above 1~GeV, where the number of 
measurements for this reaction is small.  Of particular importance in 
all energy regions is the need for data obtained involving polarized 
photons and polarized targets.  Due to the closing of hadron facilities, 
new $\pi^-p\to\gamma n$ experiments are not planned and only $\gnp$ 
measurements are possible at electromagnetic facilities using deuterium 
targets. Our agreement with existing $\pi^-$ photoproduction 
measurements leads us to believe that these photoproduction measurements 
are reliable despite the necessity of using a deuterium target. We hope 
that new CLAS $\Sigma$-beam asymmetry measurements for $\vec{\gamma}n\to
\pi^-p$, at E$_\gamma$ = 910 up to 2400~MeV and pion production angles 
from 20$^\circ$ to 140$^\circ$ (1200 data) in the CM frame, will soon 
\cite{dar} provide further constraints for the neutron multipoles.\\

\vspace{0.5in}
\begin{acknowledgments}
The authors acknowledge helpful comments and preliminary fits from 
R.~A.~Arndt in the early stages of this work.  Then the authors are 
thankful to E.~Pasyuk for useful discussions. We acknowledge the 
outstanding efforts of the CLAS Collaboration who made the experiment 
possible. This work was supported in part by the U.S. Department of 
Energy Grants DE--FG02--99ER41110 and DE--FG02---03ER41231, by the Russian 
RFBR Grant No.~02--0216465, by the Russian Atomic Energy Corporation 
``Rosatom" Grant NSb--4172.2010.2, the National Science Foundation, and 
the Italian Istituto Nazionale di Fisica Nucleare.
\end{acknowledgments}


\end{document}